%% file: avgasf.tex
\documentclass[12pt]{article}
\pdfoutput=1
\usepackage[pdftex]{graphics}
\usepackage{graphicx}
\usepackage{hyperref}
\usepackage{amssymb}
\usepackage{ulem}
\usepackage[square, comma, numbers, sort&compress]{natbib}
\usepackage{caption,subfigure}
\subcapfont{\bf}
\renewcommand{\figurename}{{\bf Figure}} 
\renewcommand{\tablename}{{\bf Table}} 
\makeatletter
\renewcommand{\fnum@figure}[1]{\textbf{\figurename~\thefigure}:}
\renewcommand{\fnum@table}[1]{\textbf{\tablename~\thetable}:}

\makeatother 

\newcommand\pubnumber{SLAC--PUB--14851 \\ SU--ITP--2012/02}
\newcommand\pubdate{}

\def\SLAC{SLAC, Menlo Park, CA 94025 \\ SITP, Stanford University,
Stanford, CA 94305}

\def\emailone{\footnote{janko@stanford.edu}}                     
\def\emailtwo{\footnote{larkoski@stanford.edu}}           

\def\Title#1{\begin{center} {\Large #1 } \end{center}}
\def\Author#1{\begin{center}{ \sc #1} \end{center}}
\def\Address#1{\begin{center}{ \it #1} \end{center}}

\newcommand\pubblock{\rightline{\begin{tabular}{l} \pubnumber\\
         \pubdate \end{tabular}}}
\newenvironment{Abstract}{\begin{quotation} \begin{center}
                       ABSTRACT
     \end{center}\bigskip  }{\end{quotation}}

\def\Acknowledgements{\bigskip  \bigskip \begin{center} \begin{large}
             \bf ACKNOWLEDGEMENTS \end{large}\end{center}}

\input mymacros.tex

\def\apb#1#2#3{\langle #1  #2  #3 ]}

\begin{document}
\begin{titlepage}
\pubblock

\vfill
\Title{Angular Scaling in Jets}
\vfill
\Author{Martin Jankowiak\emailone and Andrew J. Larkoski\emailtwo}
\Address{\SLAC}
\vfill
\begin{Abstract}
 
 We introduce a jet shape observable defined for an ensemble of jets in terms of two-particle angular correlations and a resolution parameter $R$. 
This quantity is infrared and collinear safe and can be interpreted as a scaling exponent for the angular distribution of mass inside the jet.
For small $R$ it is close to the value 2 as a consequence of the approximately scale invariant QCD dynamics.  For
large $R$ it is sensitive to non-perturbative effects.  We describe the use of this correlation function for tests
of QCD, for studying underlying event and pile-up effects, and for tuning Monte Carlo event generators.

\end{Abstract}
\vfill
\vfill
\end{titlepage}
\def\thefootnote{\fnsymbol{footnote}}
\setcounter{footnote}{1}
%
\newpage
\setcounter{page}{1}

\section{Introduction}

Over the past years, experimental and theoretical advances have made it possible
to ask increasingly detailed questions about jets.  Recently, there has been considerable interest in jet 
substructure \cite{arXiv:1012.5412}.  Much work has
focused on identifying jets initiated by boosted heavy objects such as the Higgs boson and the top quark.  A variety of techniques have been
proposed to identify and characterize substructure on a jet-by-jet basis.  It is natural to draw upon these techniques to motivate 
interesting observables for the study of QCD.  In this paper we explore an observable defined on {\it ensembles} of jets. 

A natural candidate for such an observable is the ensemble average of the angular structure function introduced in Ref.~\citep{notrees}.
We will see that this ensemble average has a clear physical interpretation in terms of an average scaling exponent. Its
leading order behavior can be found from a napkin-sized computation. 
Below we will argue that this ensemble average provides an interesting observable for at least three different reasons.  
First, it is an infrared and collinear safe observable that, in the perturbative regime, measures the extent to which QCD jet dynamics is scale invariant.
Second, because it is formulated in terms of two-particle correlations, it asks a particularly detailed question about jet substructure.  
We find that different Monte Carlo event generators give significantly
different predictions.  Consequently measuring this ensemble average could give valuable feedback on the performance of the Monte Carlo,
with any disagreements pointing towards the need for additional tuning or improvement of the physics modeling.
Finally, the ensemble average has a simple dependence on uncorrelated radiation, such as might be
expected from underlying event (UE) or pile-up (PU) contributions to jets. 
 This suggests that measurements of the ensemble average could yield
useful information about the average contribution of the underlying event and pile-up to hard perturbative jets.
More broadly, contributions to the ensemble average with different scaling behaviors will be
more or less important at smaller or larger angular scales.  In this sense the ensemble average exhibits a characteristic
sensitivity to both perturbative and non-perturbative physics, with the former dominating at small angular scales and the latter
becoming important at large angular scales.
   
The outline of this paper is as follows.  In Sec.~\ref{avgasf} we review the definition of the angular structure function and introduce
its ensemble average. In Sec.~\ref{analytics} we compute the leading order behavior of the ensemble average in the collinear
approximation and discuss
expectations for corrections to the leading order result. In Sec.~\ref{ueinjet} we investigate the sensitivity of the ensemble average
to the underlying event and pile-up, formulating a procedure for measuring the average density of uncorrelated radiation for a given
ensemble of jets. In Sec.~\ref{trans} we gain additional insight into the physics of the ensemble average by considering ensembles
of soft radiation in the transverse regions of the detector.  In Sec.~\ref{discuss} we discuss our results and present our conclusions.

\section{Average angular structure function}
\label{avgasf}

It has long been appreciated that jets have a fractal-like structure.  This point of view emerges naturally from the description
of the parton shower as a probabilistic Markov chain.  The authors of Ref.~\cite{LU-TP-91-5} have computed the fractal dimension
of a jet, while Ref.~\cite{SLAC-PUB-5593} advocates the use of the fractal phase space introduced in Ref.~\cite{Andersson:1988ee} 
as a useful diagnostic tool for complex events.  

Correlation functions provide a convenient language for studying fractal systems.  In particular they can be used to define fractal dimensions
through their limiting behavior at small scales.  With this in mind, let us review the pair of correlation functions introduced in Ref.~\cite{notrees}.  The
first is the `angular correlation function,' defined as\footnote{In Ref.~\cite{notrees} $\mathcal{G}(R)$ is normalized so that $\mathcal{G}(R)\to 1$ at large $R$; however, for the purposes
of this paper it is convenient to leave $\mathcal{G}(R)$ unnormalized with dimensions of mass squared.}
\beq
\mathcal{G}(R)\equiv\sum\limits_{i\ne j}p_{T i}p_{T j}
\Delta R_{ij}^2\Theta (R-\Delta R_{ij})
\eeq{acf}
where the sum runs over all pairs of constituents of a given jet and $\Theta(x)$ is the Heaviside step function.  Here $p_{T i}$ is the
transverse momentum of constituent $i$, and $\Delta R_{ij}$
is the Euclidean distance between $i$ and $j$ in the pseudorapidity ($\eta$) and azimuthal angle
 ($\phi$) plane: $\Delta R_{ij}^2=(\eta_i-\eta_j)^2+(\phi_i-\phi_j)^2.$ 
Infrared and collinear safety and $z$-boost invariance fix this as the unique form for a two-particle angular 
correlation function defined on the constituents of a jet, 
with the only remaining freedom being that the exponent of $\Delta R_{ij}$ in Eq.~\ref{acf} is arbitrary so long as it is positive.
$\mathcal{G}(R)$ is the contribution to a jet's mass from constituents
separated by an angular distance of $R$ or less.  It is worth emphasizing that $R$ does {\it not}
mark the distance with respect to any fixed center.  

In the context of fractals, a correlation function $c(R)$ gives rise to a corresponding correlation dimension ${\cal D}_c$
defined as \cite{fractals}:
\beq
{\cal D}_c \equiv \lim_{R\to 0}\frac{\log c(R)}{\log R}
\eeq{corrdim}
There is of course an immediate obstacle to using Eq.~\ref{acf} and Eq.~\ref{corrdim} to define the correlation dimension of a jet: the
finite resolution of the detector makes the small $R$ limit inaccessible.  In addition, the fractal-like structure of a jet
does not continue down to arbitrarily small scales, since, for a jet with transverse momentum $p_T$, the parton shower is cutoff at
an angular scale $R_{\rm{min}} \gtrsim \Lambda_{\rm{QCD}}/p_T$.  A sensible alternative to Eq.~\ref{corrdim} is to instead define 
an `angular structure function' $\Delta \mathcal{G}(R)$ via a logarithmic derivative:
\beq
\Delta {\cal G}  \equiv {d \log {\cal G} \over d \log R}={R \over {\cal G}} {d {\cal G} \over d R}= R {
\sum\limits_{i\ne j}p_{T i}p_{T j}
\Delta R_{ij}^2\delta (R-\Delta R_{ij})
\over \sum\limits_{i\ne j}p_{T i}p_{T j}
\Delta R_{ij}^2\Theta (R-\Delta R_{ij})}
\eeq{asf}
For a jet with a finite number of constituents, the $\delta$-function in Eq.~\ref{asf} results in a noisy
function of $R$.  A convenient way to obtain a smooth version of $\Delta \mathcal{G}(R)$ is to replace
the $\delta$-function by a gaussian with a fixed width $dR$:
\beq
\delta (x) \to \delta_{\rm dR}(x)={\exp\left({{-x^2 / 2{dR}^2}} \right) \over \sqrt{2\pi} dR}
\eeq{kernel}
In order to maintain the leftmost equivalence in Eq.~\ref{asf} the $\Theta$-function must also be replaced
by an error function $\Theta_{\rm dR}$ with the same width $dR$.  

The angular structure function $\Delta {\cal G}(R)$ encodes the scaling of the angular correlation function at a particular
value of $R$. In particular if ${\cal G}(R) \sim R^\beta$ then $\Delta {\cal G}(R)=\beta$.  In this sense $\Delta {\cal G}(R)$
recovers a scaling exponent analogous to the correlation dimension in Eq.~\ref{corrdim}.  On a jet-by-jet basis $\Delta {\cal G}(R)$
exhibits dramatic peaks at prominent angular scales corresponding to separations between hard substructure in the jet.  This
property of $\Delta {\cal G}(R)$ is exploited in Ref.~\cite{notrees} to construct an efficient top tagging algorithm.  In order to clearly observe scaling 
exponents, however, we will need to average over large ensembles of jets, since the number of final state particles in a single
jet is too few to clearly observe fractal structure.  Such an ensemble average will be the subject of the rest of this paper.

We use the angular correlation function as our basic object, defining its ensemble average as
\beq
\langle \mathcal{G}(R) \rangle \equiv {1 \over N} \sum_{k = 1}^N \mathcal{G}(R)_k \ 
\eeq{aveg}
where $N$ is the size of the ensemble and $\mathcal{G}(R)_k$ is the angular correlation
function of the $k$th jet.  From this average, we define the average angular structure
function:
\begin{eqnarray}
\langle \Delta \mathcal{G}(R) \rangle &\equiv& \frac{R}{\langle \mathcal{G}(R) \rangle} \frac{d}{dR} \langle \mathcal{G}(R) \rangle \CR
&=& R {  \sum_{k = 1}^N {\mathcal{G}'}(R)_k \over  \sum_{k = 1}^N \mathcal{G}(R)_k} \CR
&=& R \frac{  \sum_{k = 1}^N \sum_{i \ne j} p_{T k,i} p_{T k,j}
\Delta R_{ij}^2\delta_{\rm dR} (R-\Delta R_{ij}) }{  \sum_{k = 1}^N \sum_{i \ne j} p_{T k,i} p_{T k,j}
\Delta R_{ij}^2\Theta_{\rm dR} (R-\Delta R_{ij}) } \CR
\label{avedg}
\end{eqnarray}
where $\delta_{\rm dR}(R)$ and $\Theta_{\rm dR}(R)$ are the gaussian and error functions with width $dR$, respectively.  
Note that the ensemble average is {\it not} defined as an average over $N$ angular structure functions $\Delta\mathcal{G}(R)_k$:
\beq
\langle \Delta\mathcal{G}(R) \rangle \ne \frac{1}{N} \sum_{k = 1}^N \Delta\mathcal{G}(R)_k
\eeq{notthis}
We make this choice for at least two reasons. First, the definition in Eq.~\ref{avedg} lends itself more easily to 
analytical computation and makes possible its interpretation as an average scaling exponent.  Second, on an event-by-event
basis, the angular structure function is quite noisy.  Consequently, the ensemble average in Eq.~\ref{notthis} is significantly noisier
than that in Eq.~\ref{avedg}. 

Throughout this paper we will set $dR=0.04$.  Although nonzero $dR$ sculpts the ensemble averages somewhat,
especially near $R=0$, for $dR=0.04$ the effect is small enough that we
need not consider it when calculating $\langle{\mathcal{G}}(R)\rangle$ analytically.  

\section{Calculating the average}\label{analytics}

As we will now show, a striking property of $\langle \Delta \mathcal{G}(R) \rangle$ is that its leading order
behavior can be understood from a simple computation.  This is in contrast to, e.g., the integrated jet shape
$\Psi(R)$ \cite{Ellis:1992qq}, which requires a detailed calculation even for its leading order behavior \cite{hep-ph/9707338}.  
This section is organized as follows.  In Sec.~\ref{dla} we compute
the leading order behavior of $\langle \Delta \mathcal{G}(R) \rangle$ in the collinear
approximation.  While it would be rewarding to perform a NLO computation 
of $\langle \Delta \mathcal{G}(R) \rangle$, in the subsequent sections we limit ourselves to exploring some of the features we expect to 
emerge from a more complete calculation. This task will be made easier
thanks to the clear physical interpretation of $\langle \Delta \mathcal{G}(R) \rangle$ as a scaling exponent. 
First, in Sec.~\ref{running} we discuss the qualitative effect of the running of the strong coupling.   
Second, in Sec.~\ref{higherorder} we explore higher order effects with an emphasis on the expected difference between quark and gluon jets.
Finally, in Sec.~\ref{factorization} we briefly touch upon whether $\langle \Delta \mathcal{G}(R) \rangle$ could be amenable to factorization.

\subsection{Collinear approximation}\label{dla}

To begin we compute the average value of the angular correlation function
$\langle {\cal G}(R) \rangle$ in the collinear approximation.  
To first order in $\alpha_s$, $\langle {\cal G}(R) \rangle$ can be computed from
\beq
\langle {\cal G}(R) \rangle \simeq \frac{\alpha_s}{2\pi} p_T^2\int^{R_0^2}_0 \frac{d\theta^2}{\theta^2}
\int_0^1 dz \ P(z)  z(1-z)\theta^2 \Theta(R-\theta)
\eeq{avega}
where $R_0$ is the radius of the jet algorithm and $P(z)$ is the appropriate Altarelli-Parisi splitting function. 
Notice that, as discussed at the end of Sec.~\ref{avgasf}, for the purposes of this section it is enough to
set $dR=0$, although $dR>0$ will be needed for any actual measurement.
Since we are interested in the interior of the jet, in all of the following expressions we will assume that
$R<R_0$.  This prevents us from making predictions about 
edge effects in $\langle {\cal G}(R) \rangle$, but we do not expect the collinear approximation
to be a good approximation at larger $R$ anyway.   From Eq.~\ref{avega} we find:
\beq
\langle {\cal G}(R) \rangle = \frac{\alpha_s}{2\pi} p_T^2 R^2 
\left\{
\begin{array}{lr}
\frac{3}{4} C_F & \;\;\;\;\rm{quark\ jets}\\ \\
\frac{7}{10}C_A+\frac{1}{10}n_F T_R & \;\;\;\;\rm{gluon\ jets}
\end{array}
\right.
\eeq{qga}
We thus have the leading order result that the angular correlation function for QCD jets
goes like $R^2$.  
Consequently, for both quark and gluon jets, we have that $\langle \Delta {\cal G}(R) \rangle = 2$.  
Note that this holds for any jet algorithm.

Some interpretation of this result is in order.  The average angular structure function $\langle \Delta {\cal G}(R) \rangle$ is
a measure of how energy is distributed within a jet.  A typical QCD jet has a hard core with 
the structure of emissions in and around the core controlled by the soft and collinear singularities.
The fact that for QCD jets $\langle \Delta {\cal G}(R) \rangle =2 $ at leading order tells us that QCD has a collinear
singularity of strength $d\Theta^2 / \Theta^2$.  By contrast, if the energy were distributed uniformly over the entire cone of the jet, we would expect $\langle {\cal G}(R) \rangle \sim R^4 $ and $\langle \Delta {\cal G}(R) \rangle \simeq 4$, at least up to edge effects.

\subsection{Running coupling}\label{running}
The qualitative effects of including a running coupling are straightforward to understand.  We proceed by evaluating the strong coupling
at a characteristic energy scale $Q^2 = p_T^2 \theta^2 g(z)$, where $g(z)$ is a function of the momentum fraction $z$ that is regular for 
$0\leq z \leq 1$.  Typically, $Q$ is taken to be the jet mass ($g(z) = z(1-z)$) or  
transverse momentum of the emitted gluon ($g(z)=1-z$), 
but for our purposes what is most important is that $Q^2$ is proportional to $\theta^2$.  
This implies that at small angles the strong coupling increases resulting in more radiation in the small angle region of phase space.  
Effectively, this increases the strength of the collinear singularity with respect to the fixed coupling expectation.  Thus, we expect that the 
effect of a running coupling is to 
lower the value of 
$\langle \Delta {\cal G}(R) \rangle$ with respect to the fixed coupling result.

A simple calculation confirms this picture.  Now including a running coupling we have:
\beq
\langle {\cal G}(R) \rangle \simeq p_T^2 \int \frac{d\theta^2}{\theta^2}
\int dz \ P(z)  z(1-z) \frac{\alpha_s(p_T^2 \theta^2 g(z))}{2\pi} \theta^2 \Theta(R-\theta) \ 
\eeq{avegarun}
To lowest order the running coupling is
\beq
\alpha_s(Q^2) = {\alpha_0 \over \log\left( {p_T^2 \theta^2 g(z) \over \Lambda^2_{\rm{QCD}} }\right)} \ 
\eeq{alpharun}
where $\alpha_0 = 2\pi / \beta_0$ and $\beta_0$ is the leading coefficient of the QCD beta function.  
In the limit that $p_T R \gg \Lambda_{\rm{QCD}}$, the precise forms of $g(z)$ and $P(z)$ are irrelevant, and we find that
\beq
\langle {\cal G}(R) \rangle \propto \frac{R^2}{\log \left( {p_T R \over \Lambda_{\rm{QCD}}} \right)}
\eeq{gapprox}
with the result that
\beq
\langle \Delta {\cal G}(R) \rangle 
\simeq 2- {1 \over  \log \left( {p_T R \over \Lambda_{\rm{QCD}}} \right)}
\eeq{dgapprox}
As expected, including the running coupling decreases the average angular structure function.  
To first order in $1 / \log \left( {p_T R \over \Lambda_{\rm{QCD}}} \right)$,
Eq.~\ref{dgapprox} is true for both quark and gluon jets.  This effect is not negligible.  For example, for $p_T = 200 \rm{\;GeV}$,
 $R=1.0$, and $\Lambda_{\rm{QCD}}=300 \rm{\;MeV}$,
we have that $1 /  \log \left( {p_T R \over \Lambda_{\rm{QCD}}} \right) \simeq 0.15$.

\subsection{Higher order effects}\label{higherorder}

\begin{figure}
\centering
    \includegraphics[width=7.1cm]{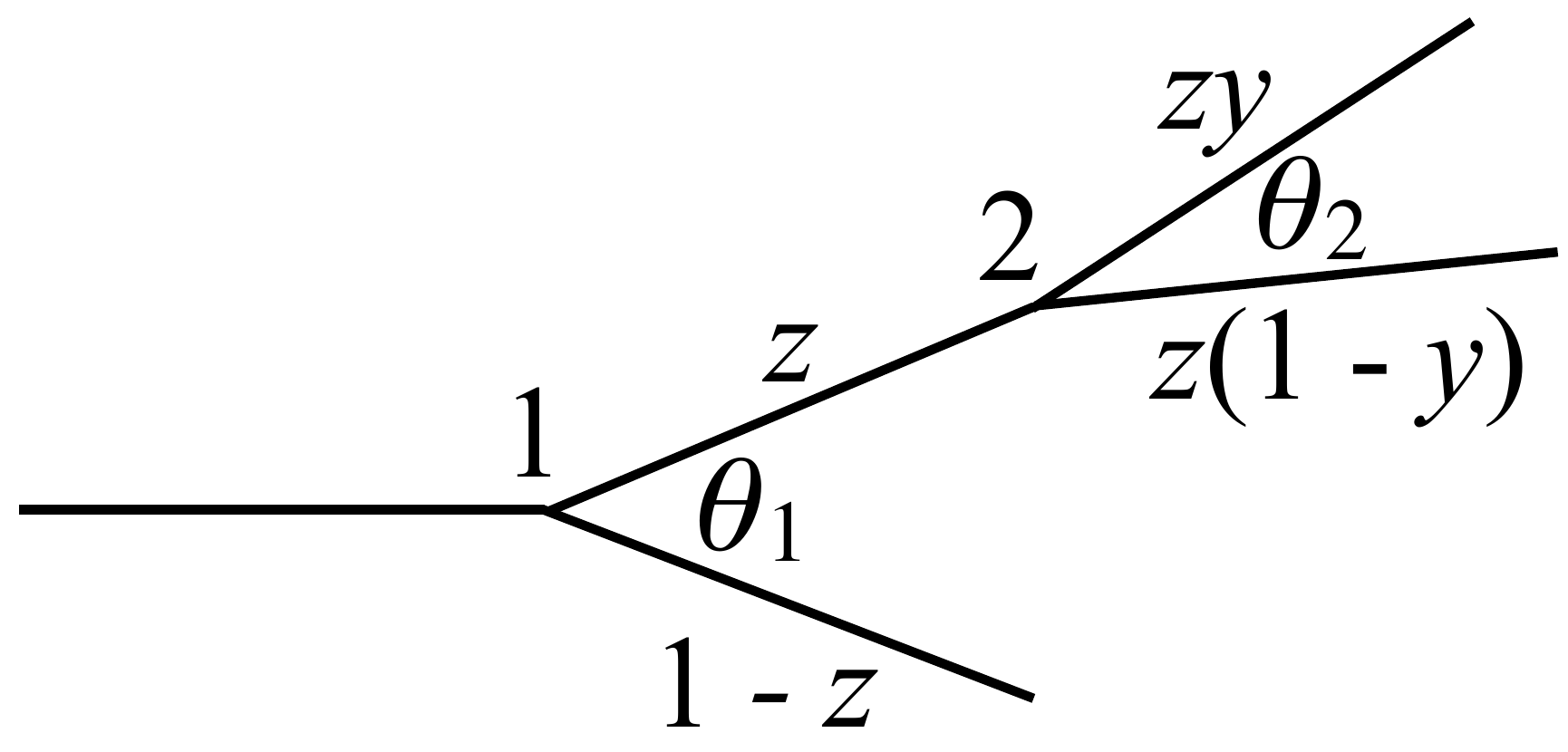}
\caption{Splitting diagram for the ${\cal O}(\alpha_s^2)$ contribution to $\langle {\cal G}(R) \rangle$.}\label{aakin}
\end{figure}

We can get an idea of the nature of the higher order corrections
to the ${\cal O}(\alpha_s)$ fixed coupling result by continuing the calculation of Sec.~\ref{dla} to ${\cal O}(\alpha_s^2)$.  
 The kinematic
identifications of the three parton final state are illustrated in Fig.~\ref{aakin}, which depicts the longitudinal momentum
fractions and angles associated to each splitting.  We impose angular ordering so that $\theta_1 \ge \theta_2$. 
The expression for the ${\cal O}(\alpha_s^2)$ contribution to $\langle {\cal G}(R) \rangle$ is then given by:
\begin{eqnarray}\label{avegaa}
\langle {\cal G}(R) \rangle_{\alpha_s^2}& \simeq& \biggl(  \frac{\alpha_s}{2\pi} \biggr)^2p_T^2 
\int_0^{R_0^2} \frac{d\theta_1^2}{\theta^2_1} \int_0^{\theta_1^2} \frac{d\theta_2^2}{\theta^2_2}
\int_0^1 dz \int_0^1 dy\ P_1(z) P_2(y) \CR
& & \biggl\{
z^2 y(1-y)\theta^2_2 \Theta(R-\theta_2) \CR
&& + z(1-z)y \int_0^{2\pi} \frac{d\phi}{2\pi}\biggl[ \biggl( (1-y)^2\theta_2^2+\theta_1^2-2(1-y)\theta_1\theta_2\cos\phi\biggr) \CR
&& \ \ \times\Theta\biggl(R-\sqrt{(1-y)^2\theta_2^2+\theta_1^2-2(1-y)\theta_1\theta_2\cos\phi}\biggr)\biggr] \CR
&&+ z(1-z)(1-y) \int_0^{2\pi} \frac{d\phi}{2\pi}\biggl[ \biggl( (y^2\theta_2^2+\theta_1^2+2y\theta_1\theta_2\cos\phi\biggr) \CR
&& \ \ \times\Theta\biggl(R-\sqrt{y^2\theta_2^2+\theta_1^2+2y\theta_1\theta_2\cos\phi}\biggr)\biggr] 
\biggr\}
\end{eqnarray}
Using unpolarized splitting functions and setting $\cos\phi = 0$ results in a simple analytic formula given by
\beq
\langle {\cal G}(R) \rangle_q = \frac{\alpha_s}{2\pi}p_T^2 R^2 \left\{
\frac{3}{4} C_F  +\frac{\alpha_s}{2\pi}
\left(
-\frac{25}{16}C_F^2+\frac{49}{120} C_A  C_F +\frac{7}{120} n_F T_R  C_F
\right)
\left(
1 + \log\frac{R_0^2}{R^2}
\right)
\right\}
\eeq{aveasfq}
for quark jets and
\begin{eqnarray}\label{aveasfg}
\langle {\cal G}(R) \rangle_g & =& \frac{\alpha_s}{2\pi}p_T^2 R^2 \left\{
\frac{7}{10}C_A + \frac{1}{10} n_F T_R \right.
\CR
&& \left.+\frac{\alpha_s}{2\pi}
\left(
-\frac{49}{100} C_A^2 -\frac{91}{300}C_A n_F T_R 
-\frac{1}{30} n_F^2 T_R^2 + \frac{7}{20}C_F n_F T_R
\right)
\left(
1+\log\frac{R_0^2}{R^2}
\right)
\right\} \CR
\end{eqnarray}
for gluon jets.  From these expressions we can calculate $\langle \Delta {\cal G}(R)\rangle$ to ${\cal O}(\alpha_s)$, finding
\begin{eqnarray}\label{quarkaasf}
\langle \Delta {\cal G}(R)\rangle_q &\simeq& 2 - \frac{\alpha_s}{\pi}\left(
-\frac{25}{12}C_F+\frac{49}{90} C_A +\frac{7}{90} n_F T_R\right)+{\cal O}(\alpha_s^2)\nonumber \\
&\simeq&2+0.95\frac{\alpha_s}{\pi} 
\end{eqnarray}
for quark jets and 
\begin{eqnarray}\label{glueaasf}
\langle \Delta {\cal G}(R)\rangle_g &\simeq& 2 - \frac{\alpha_s}{\pi}\left(\frac{-\frac{49}{10} C_A^2 -\frac{91}{30}C_A n_F T_R 
-\frac{1}{3} n_F^2 T_R^2 + \frac{7}{2}C_F n_F T_R}{7C_A+n_F T_R}\right)+{\cal O}(\alpha_s^2) \nonumber \\
&\simeq& 2+2.44\frac{\alpha_s}{\pi} 
\end{eqnarray}
for gluon jets, where we have set $n_F=5$ in evaluating the color factors.  Notice that, since
\beq
\frac{d}{d\log R} \log \left [R^2(1+\epsilon \log(R^2)) \right]= 2+2\epsilon+{\cal O}(\epsilon^2)
\eeq{logflat}
$\langle \Delta {\cal G}(R)\rangle_{q/g}$ remain flat in $R$ to this order in $\alpha_s$.  In particular
Eq.~\ref{logflat} implies that the $\alpha_s^2 R^2$ terms in Eq.~\ref{aveasfq} and Eq.~\ref{aveasfg} do not contribute
to Eq.~\ref{quarkaasf} and Eq.~\ref{glueaasf} at ${\cal O}(\alpha_s)$.  These terms {\it do} contribute to the normalization of
the angular correlation function, but we do not expect them to be correctly given by the collinear approximation.  We suspect,
however, that the $R^2 \log \frac{R_0^2}{R^2}$ terms are robust as far as the scaling exponent is concerned.

The increase of $\langle \Delta {\cal G}(R)\rangle$ with respect to the leading order result can be loosely interpreted
as resulting from a relative increase in the amount of energy radiated away from the center of the jet.  This increase is
largest for gluon jets as a result of the large associated color factors.

\begin{figure}
\centering
\subfigure[Pythia8]
{
    \includegraphics[width=6.8cm]{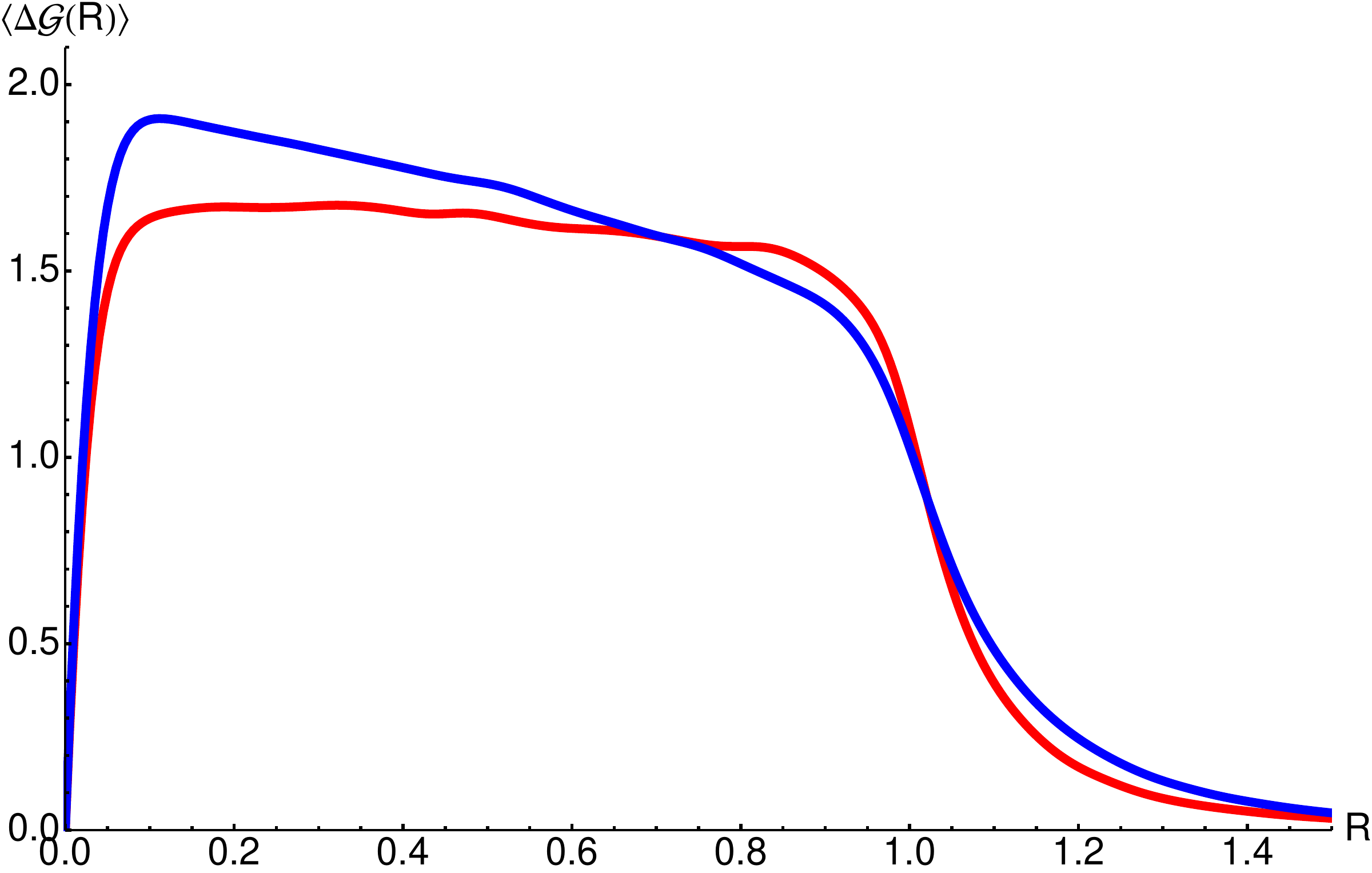}\label{pythianoue}
}
\hspace{.1cm}
\subfigure[Herwig++]
{
    \includegraphics[width=6.8cm]{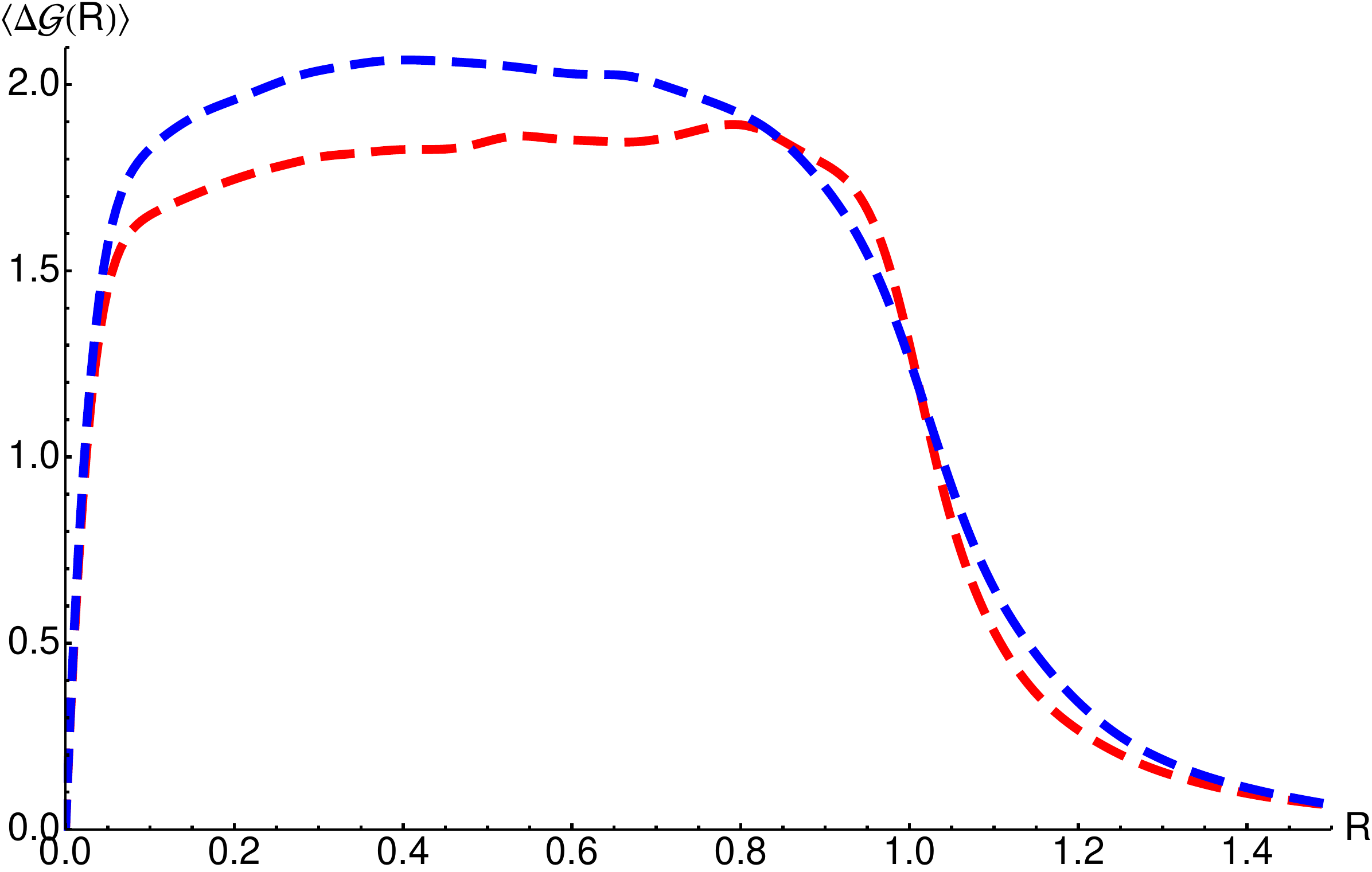}\label{herwignoue}
}
\caption{Average angular structure functions for ensembles of jets with $p_T>$ 200 GeV and no underlying event or initial state radiation.  
Red curves denote quark jets and blue curves denote gluon jets.  These are anti-kT jets with jet radius $R_0=1.0$ as generated
with Pythia8 (left) and Herwig++ (right).
See Appendix \ref{mc} for more details about the Monte Carlo.}\label{pyher_noue}
\end{figure}
\begin{figure}[h]
\centering
\subfigure[Quark jets]
{
    \includegraphics[width=6.8cm]{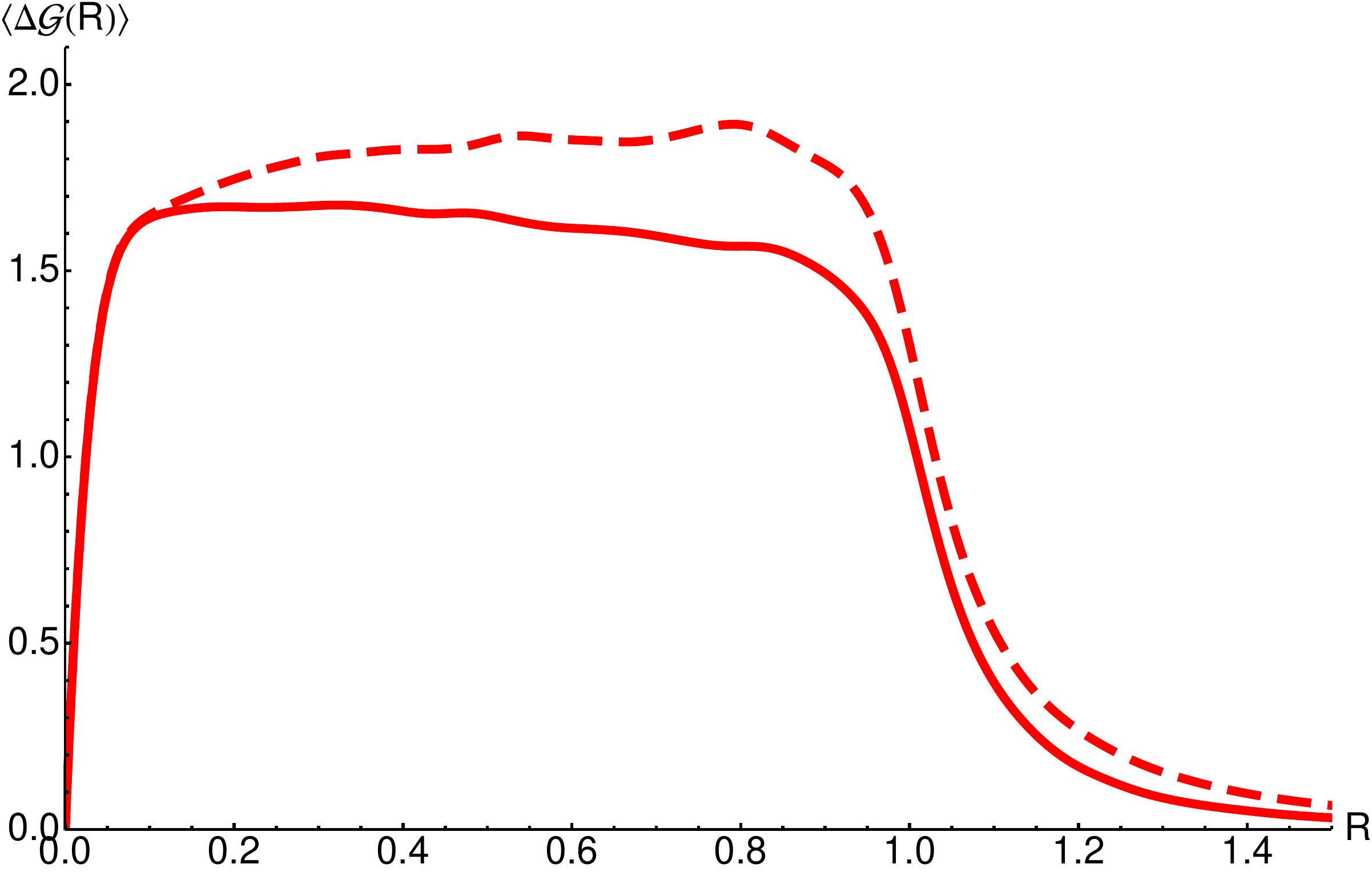}
    \label{qq_noue}
}
\subfigure[Gluon jets]
{
    \includegraphics[width=6.8cm]{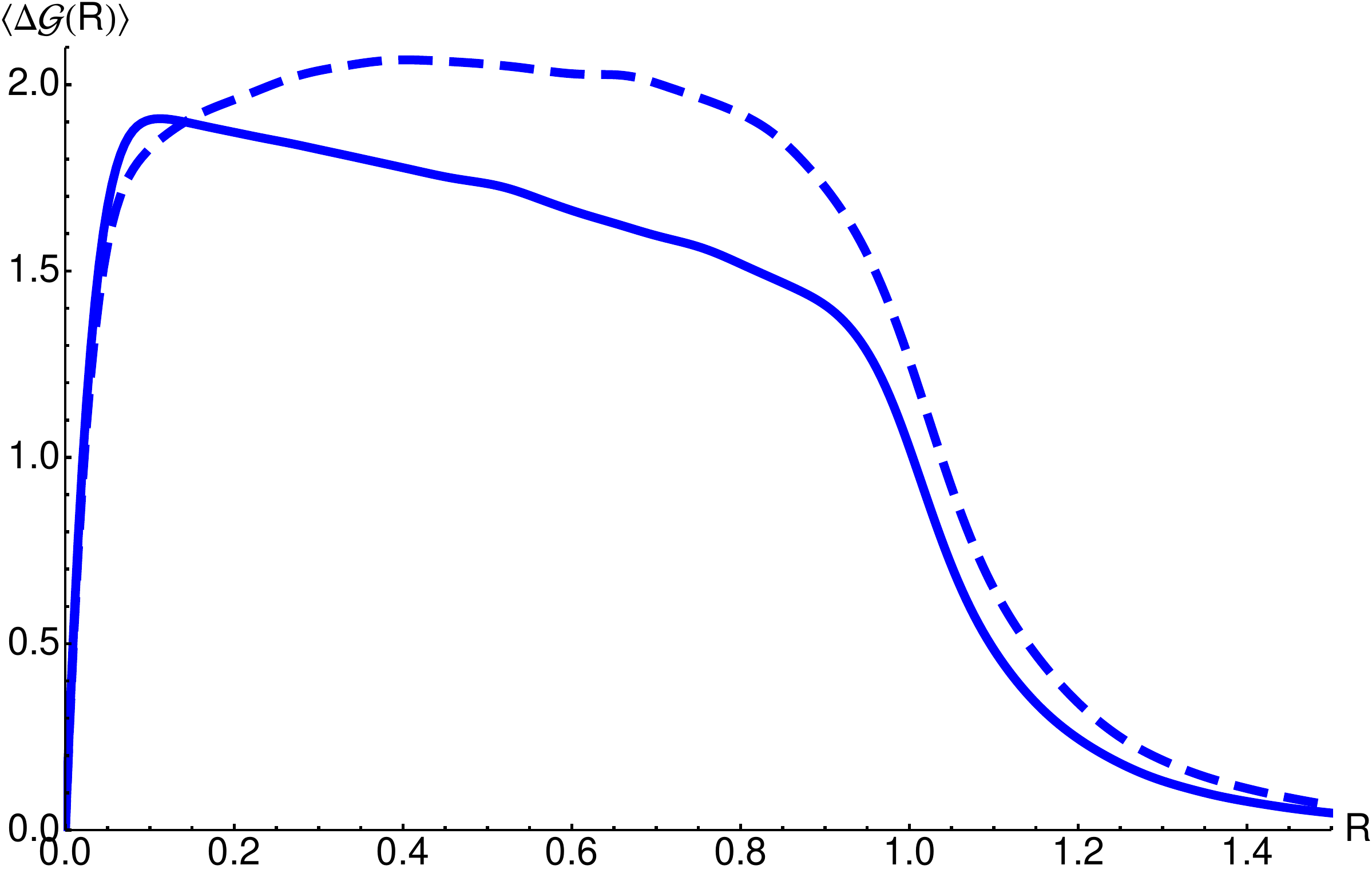}
} \\
\subfigure[Jets]
{
    \includegraphics[width=6.8cm]{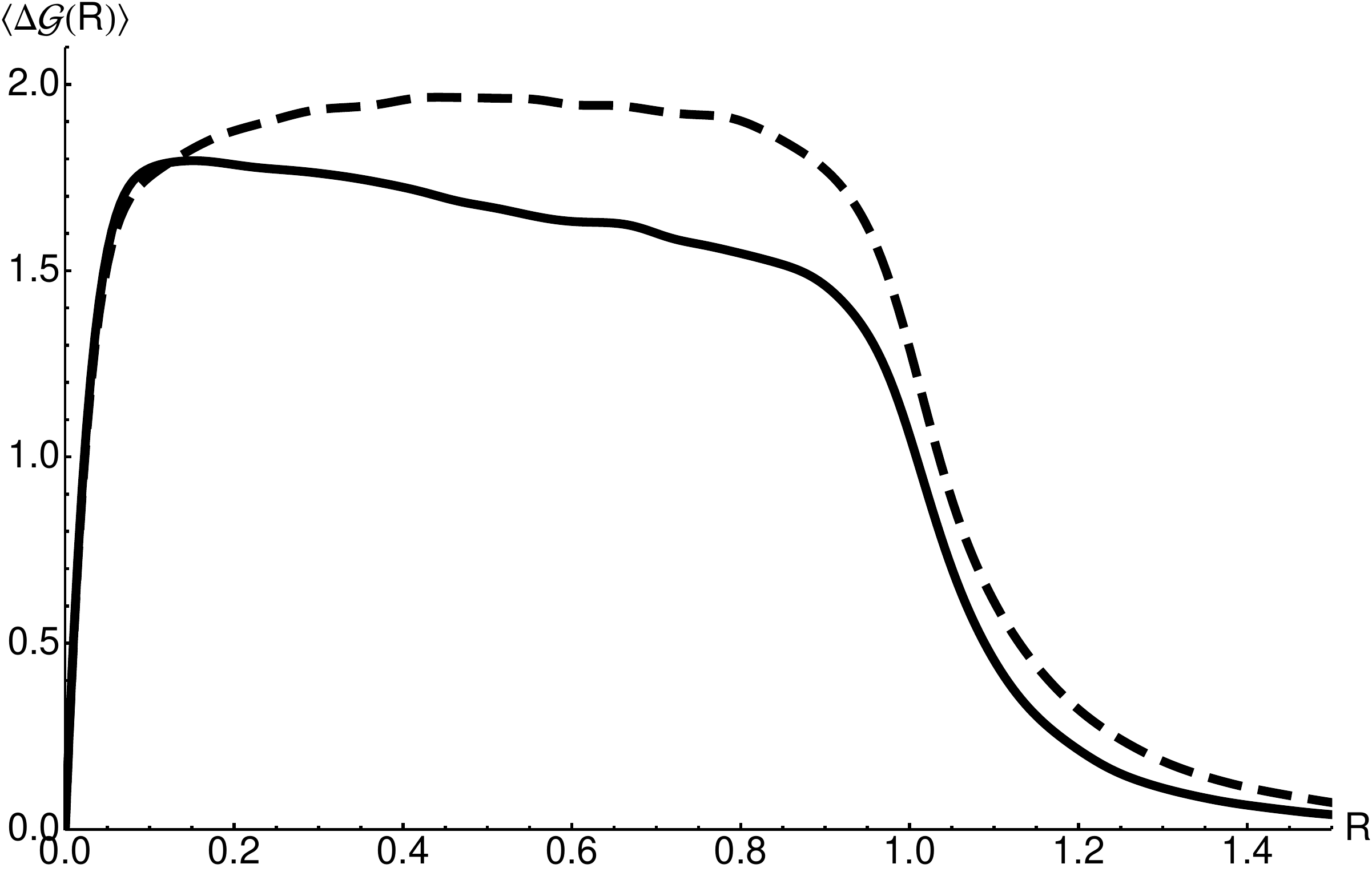}
}

\caption{Average angular structure functions for three different ensembles of jets with $p_T>$ 200 GeV and no underlying event or initial state radiation.  These are anti-kT jets with jet radius $R_0=1.0$ as generated with Pythia8 (solid) and Herwig++ (dashed). See Appendix \ref{mc} for more details about the Monte Carlo.}\label{qqgg_noue}
\end{figure}

Putting together the results from Sec.~\ref{dla}, Sec.~\ref{running}, and Sec.~\ref{higherorder} our expectations for the form
of $\langle \Delta {\cal G}(R)\rangle$ for quark and gluon jets are as follows.  Apart from edge effects, we
expect both $\langle \Delta {\cal G}(R)\rangle_q$ and $\langle \Delta {\cal G}(R)\rangle_g$ to be approximately 
flat in $R$ as in the leading order result.  Furthermore, if the effects of a running coupling are dominant then we also expect that
$\langle \Delta {\cal G}(R)\rangle_{q/g} \lesssim 2$.  In addition we expect that $\langle \Delta {\cal G}(R)\rangle_q<\langle \Delta {\cal G}(R)\rangle_g$.  
 Note that, even in the case when $\langle \Delta {\cal G}(R)\rangle_{q/g}$ are separately flat, an ensemble average over an admixture of
quark and glue jets will in general not be flat.

In Fig.~\ref{pyher_noue} and Fig.~\ref{qqgg_noue} we plot $\langle \Delta {\cal G}(R)\rangle$ for both quark and gluon jets as obtained
from Pythia8 and Herwig++ dijet events.  The ensembles are composed of anti-kT jets with $R_0=1.0$ and $p_T > 200$ GeV. Initial state radiation (ISR)
and the underlying event have been turned off.  The Monte Carlo is roughly in accord with our expectations.  Not surprisingly, our expectations
are more in line with the angular ordered shower in Herwig++, where the expected difference between quark jets and glue jets, 
i.e.~$\langle \Delta {\cal G}(R)\rangle_q<\langle \Delta {\cal G}(R)\rangle_g$, is unambiguous.  The differences between
the predictions for $\langle \Delta {\cal G}(R)\rangle$ given by Pythia8 and Herwig++ are striking (see Fig.~\ref{qqgg_noue}),
with $\langle \Delta {\cal G}(R)\rangle$
being substantially higher for Herwig++ over a large range in $R$.  This difference\footnote{ We have also
performed Monte Carlo calculations of $\langle \Delta {\cal G}(R)\rangle$ for an ensemble of jets produced in the process $e^+e^-\to q \bar{q}$.  
In this case the differences between Pythia8, Herwig++, and Pythia8+Vincia (showering done with Vincia 1.0.26 \cite{Giele:2011cb}) are small compared to 
the differences in Fig.~\ref{qq_noue}.}, which persists when the underlying event
and initial state radiation are included (see Fig.~\ref{pyherjjue}), should be measurable and motivates making the measurement.  
It would be interesting to have a detailed understanding of how this qualitatively 
different behavior emerges from the two codes; doing so, however, lies outside the scope of this paper.

\subsection{Factorization}\label{factorization}

Before moving on we would like to draw a connection to a recent analysis on the factorizability of
jet substructure observables in soft-collinear effective theory (SCET) \citep{jetfactor}.  Walsh and Zuberi determined necessary conditions for jet
substructure observables to be factorizable in the sense that any such observable can be computed as a direct product of universal terms
up to power corrections.  A central result of their analysis is that in order for factorization to hold an observable must not
demand that soft modes individually resolve collinear modes.  Because the angular correlation function is a two-particle correlation function,
one might worry that it mixes soft and collinear modes in a way that upsets factorization.  However, to leading power soft modes contributing
to $\langle {\cal G}(R) \rangle$ do not resolve individual collinear modes:
\begin{equation}\label{acfexpan}
\sum_{C,S} p_{TC}p_{TS}\Delta R^2_{CS}\Theta(R-\Delta R_{CS}) = p_{T\rm{J}}\sum_{S} p_{TS}\Delta R^2_{S\rm{J}}\Theta(R-\Delta R_{S\rm{J}}) 
\end{equation}
Here the sum runs over collinear ($C$) and soft ($S$) modes and $J$ refers to the jet.
This suggests that the ensemble average of the angular correlation function should be factorizable in SCET.

\section{Effect of uncorrelated radiation}\label{ueinjet}

In the previous section we discussed the shape of $\langle \Delta {\cal G}(R) \rangle$ as determined by the perturbative
final state shower.  At a hadron collider the colored initial state means that the dynamics of jets cannot be
understood separately from the underlying event and initial state radiation.
For convenience in the following we will often collectively refer to any radiation that is not associated with the hard, 
perturbative final state as the ``underlying event.''  In particular what we have in mind is comparably soft radiation 
that is uncorrelated with the hard scatter and which is approximately uniformly distributed in pseudorapidity and 
azimuth with transverse momentum density $\Lambda_{\rm UE}$. 

\subsection{Background}\label{uebackground}
\begin{figure}
\centering
\subfigure[Pythia8]
{
    \includegraphics[width=6.8cm]{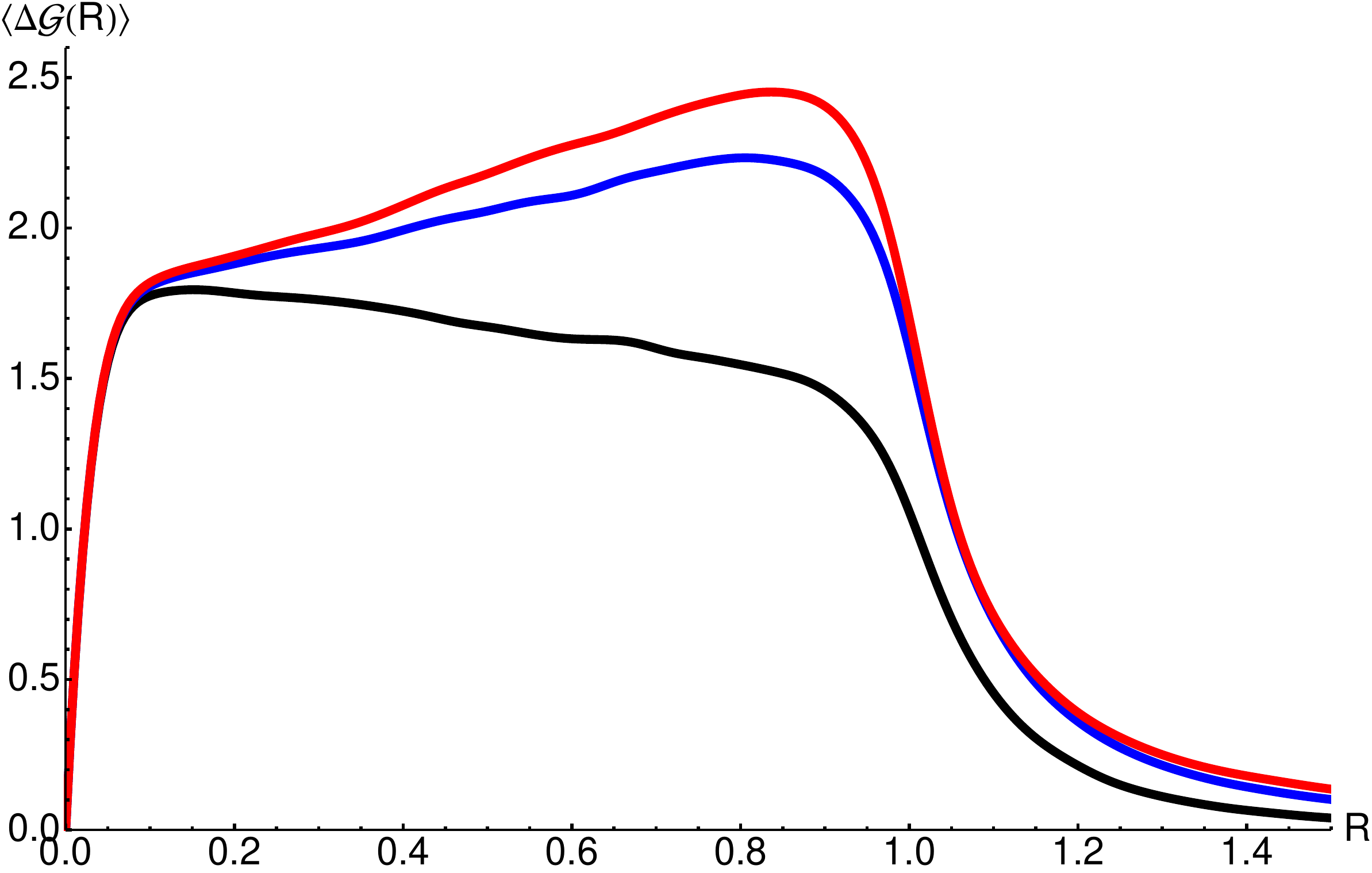}
}
\hspace{.1cm}
\subfigure[Herwig++]
{
    \includegraphics[width=6.8cm]{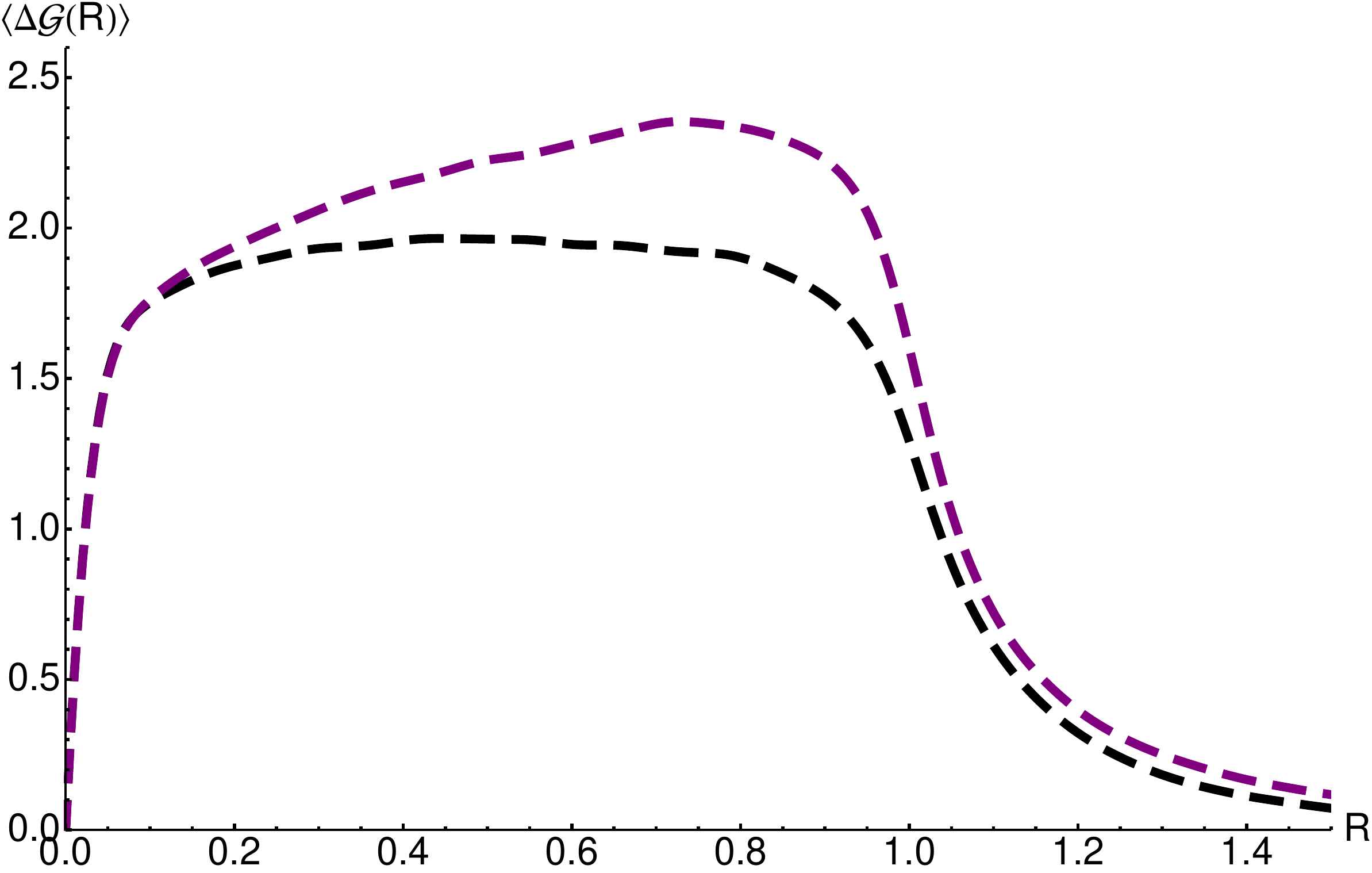}
}
\subfigure[Pythia8 vs. Herwig++]
{
    \includegraphics[width=6.8cm]{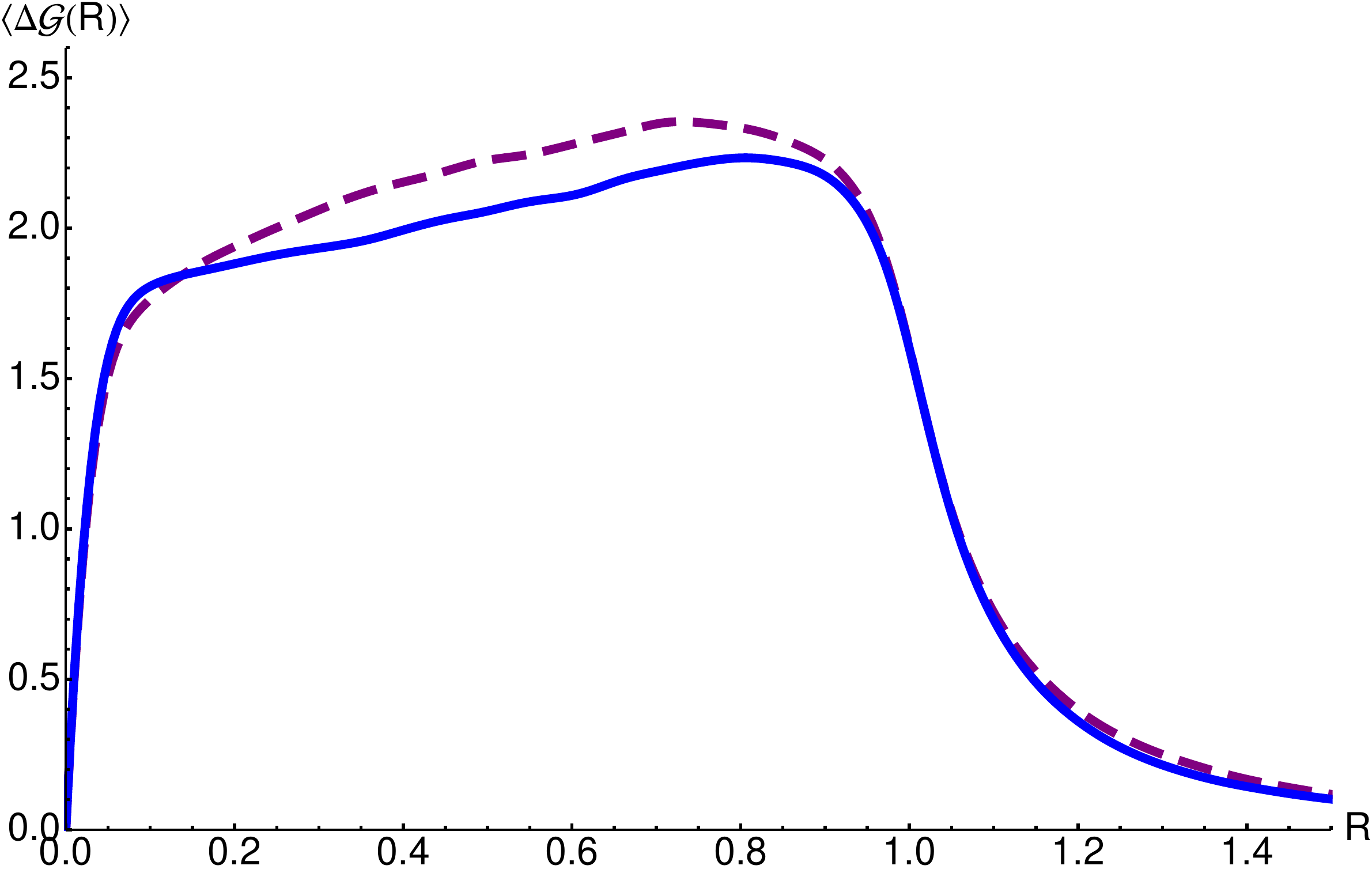}\label{pyherjjue}
}
\caption{Average angular structure functions for ensembles of jets with $p_T>$ 200 GeV.  The black curves have no underlying event nor any ISR.
In contrast to the previous section, here underlying event and ISR are turned on for the colored curves.  On the LHS, the Pythia8 samples
make use of tune 4C, with the red curve 
having twice as much UE activity as the blue curve.  On the RHS, the purple curve corresponds to Herwig++ tune LHC-UE7-2.
For comparison, the bottom figure overlays the Pythia8 and Herwig++ curves.  These are anti-kT jets with jet radius $R_0=1.0$.  See Appendix \ref{mc} for more details about the Monte Carlo.}\label{pyher_ue}
\end{figure}

We would like to ask how this UE affects $\langle \Delta {\cal G}(R) \rangle$ for an ensemble of jets of a given $p_T$. Since
by assumption $\Lambda_{\rm UE} \ll p_T$, we can neglect correlations of order $\Lambda_{\rm UE}^2$.
Furthermore, since most of the energy of the final state shower is localized in a hard core at the center
of the jet, we can write the ${\cal O}(\Lambda_{\rm UE})$ contribution to $\langle {\cal G}(R) \rangle$ as
\beq
\langle {\cal G}(R) \rangle_{\Lambda_{\rm UE}}=p_T \Lambda_{\rm UE} \int_0^{2\pi} d\phi \int_0^R R' dR' R'^2 
=\frac{\pi}{2}p_T \Lambda_{\rm UE} R^4 \ 
\eeq{UEpertcorr}
 This ansatz neglects edge effects due to the finite size of the jet, but
these are expected to be small away from $R=R_0$ for the approximately circular\footnote{See, e.g., Fig. 7 in Ref.~\cite{Salam:2009jx}
for `typical' jet shapes as generated by different algorithms.} anti-kT jets used throughout this paper.  The
range of $R$ for which this ansatz is valid will be smaller for other sequential jet algorithms (e.g.~Cambridge/Aachen and $k_T$)
that yield more irregularly shaped jets.  Thus we have:
\beq
\langle {\cal G}(R)\rangle = \langle {\cal G}(R)\rangle_{\rm pert}+\frac{\pi}{2}p_T \Lambda_{\rm UE} R^4 \ 
\eeq{acfwUE}
Since the perturbative piece goes approximately like $R^2$, at large $R$ the UE contribution is increasingly
important.  In the absence of the perturbative piece we would have $\langle {\Delta \cal G}(R)\rangle\simeq 4$.  Thus
the inclusion of UE has the effect of increasing $\langle \Delta {\cal G}(R)\rangle$ from its perturbative
value of near $2$ at small $R$ towards the value of $4$ characteristic of uniform radiation at large $R$.  This behavior 
is evident in Fig.~\ref{pyher_ue}.
Going to the average angular structure function, we can rewrite Eq.~\ref{acfwUE} as:
\beq
\label{asfwithue}
\langle \Delta {\cal G}(R)\rangle_{\rm pert}=\frac{R\frac{d}{dR}\langle {\cal G}(R)\rangle - 2\pi p_T \Lambda_{\rm UE}R^4}{\langle {\cal G}(R)\rangle - \frac{\pi}{2} p_T \Lambda_{\rm UE}R^4}
\eeq{asfpert}
Provided that we have some sort of estimate of the perturbative piece $\langle \Delta {\cal G}(R)\rangle_{\rm pert}$,
the distinctive $R$-dependence of Eq.~\ref{asfpert} can be used to fit for $\Lambda_{\rm UE}$.  In the remainder
of this section we will investigate a procedure for doing so.
 
 
\begin{figure}[t] \centering
    \includegraphics[width=9.4cm]{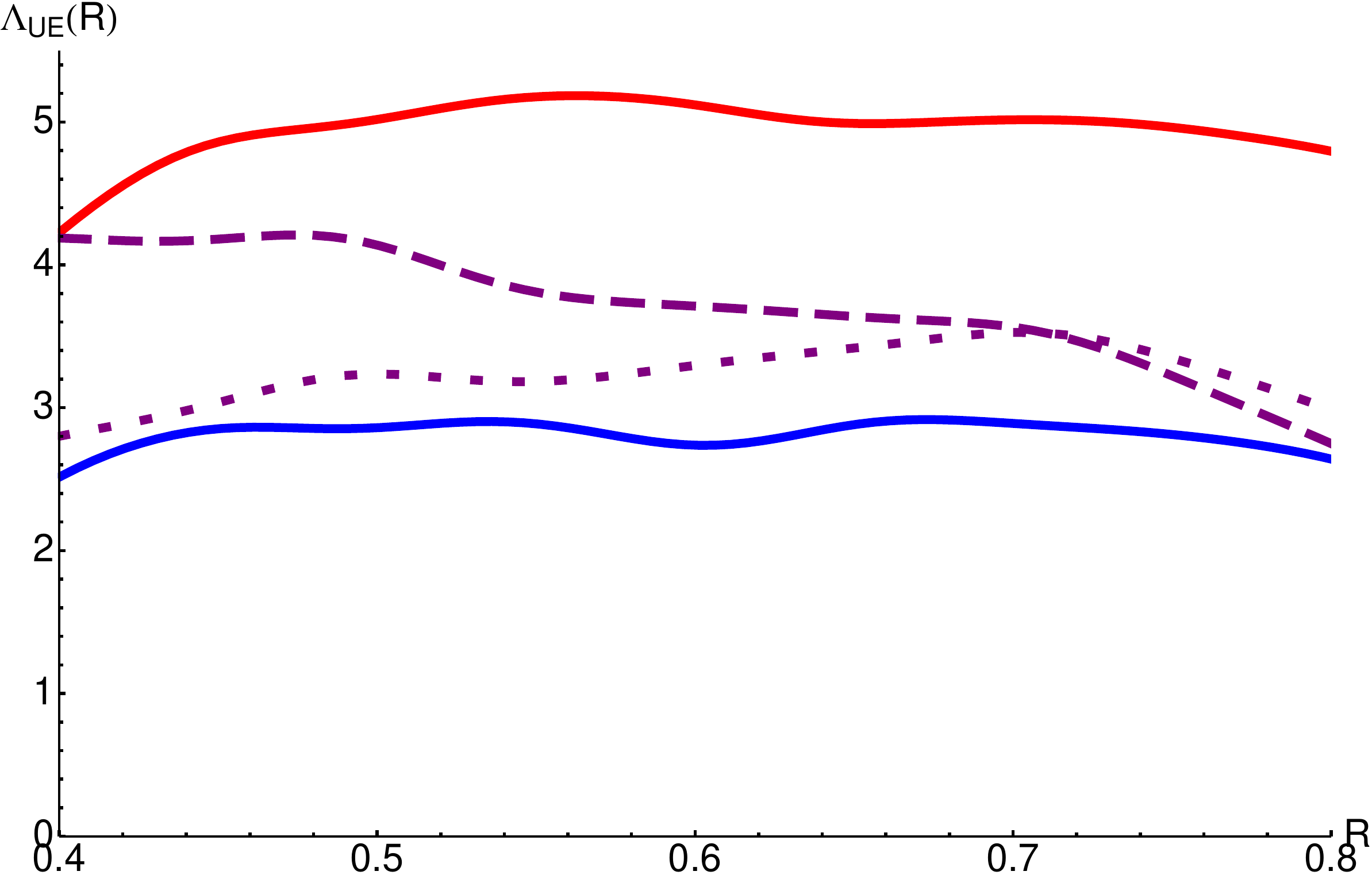}
\caption{$\Lambda_{\rm UE}(R)$ curves as extracted via the procedure detailed in Sec.~\ref{proc}.  The solid curves correspond to the Pythia8
samples in Fig.~\ref{pyher_ue}, while the dashed curve corresponds to the Herwig++ sample.  The dotted curve 
corresponds to the same Herwig++ sample as the dashed curve, the difference being that the power law of the underlying event
ansatz is changed from $R^4$ to $R^{3.3}$. The matching was done at $R_{\rm min}=0.25$.}\label{pyher_lambdacomp}
\end{figure}

\subsection{Procedure}\label{proc}
One possible procedure for extracting $\Lambda_{\rm UE}$ for a given ensemble of jets 
works from the assumption that $\langle \Delta {\cal G}(R)\rangle_{\rm pert}$ is approximately flat in $R$.  We have seen
some tentative evidence for this hypothesis in Sec.~\ref{analytics}.
We can estimate $\langle \Delta {\cal G}(R)\rangle_{\rm pert}$
by measuring $\langle \Delta {\cal G}(R)\rangle$ at small $R$.  Specifically we assume that the perturbative
piece is exactly flat and given as
\beq
\langle \Delta {\cal G}(R)\rangle_{\rm pert}= \langle \Delta {\cal G}(R_{\min})\rangle\equiv C 
\eeq{pertans}
where $R_{\min}\ll1$.  We choose $R_{\min}=0.25$.  With this ansatz, we can invert
Eq.~\ref{asfpert} to solve for $\Lambda_{\rm UE}$ as a function of $R$:
\beq
\Lambda_{\rm UE}(R)=\frac{2\langle {\cal G}(R)\rangle}{\pi p_T R^4}\frac{\langle \Delta {\cal G}(R)\rangle-C}{4-C} \ 
\eeq{lambda}
The flatness of $\Lambda_{\rm UE}(R)$ over a wide range of $R$ can then justify the ansatz {\it a posteriori}, and
 $\Lambda_{\rm UE}(R)$ can be averaged over a range in $R$ to obtain $\bar \Lambda_{\rm UE}$.  
 Summarizing, the procedure would be:
\begin{enumerate}
\item Measure $C\equiv\langle \Delta{\cal G}(R_{\min})\rangle$ and take this as an ansatz for the perturbative contribution
to $\langle \Delta{\cal G}(R)\rangle$ for all $R$.
\item Construct Eq.~\ref{lambda} for a range of $R$ with $R > R_{\min}$.
\item Then
$\bar \Lambda_{\rm UE}$, the average value of $\Lambda_{\rm UE}(R)$ 
over a range in $R$, gives a measure of the average density of UE in the ensemble of jets.
\end{enumerate}
An additional sanity check on the extracted scales could be to measure $\bar \Lambda_{\rm UE}$ as a function of  
the number of primary vertices $n_{\rm PV}$.  If the above procedure makes sense, $\bar \Lambda_{\rm UE}$
should exhibit a clear linear dependence on $n_{\rm PV}$, since pile-up will contribute uncorrelated radiation
to the ensemble of jets in an amount proportional to $n_{\rm PV}-1$. 

\subsection{Results}
Curves $\Lambda_{\rm UE}(R)$ as extracted from Monte Carlo data via this procedure are plotted in Fig.~\ref{pyher_lambdacomp}.
For Pythia8, $\Lambda_{\rm UE}(R)$ is approximately flat over a broad range in $R$, justifying the ansatz in
Eq.~\ref{acfwUE}.  Note that the scale of the red curve is about twice that of the blue curve.  This is as expected, since
the red sample has twice as much UE activity as the blue sample (but the same amount of ISR).  The dashed Herwig++ curve, by contrast, 
is not flat and has an unambiguous downward slope.  Changing the UE power law ansatz from 
$R^4$ to $R^{3.3}$ results in a $\Lambda_{\rm UE}(R)$ curve that is much flatter, though still at a similar scale.  That is to say that
for Herwig++ the underlying event is not uniformly distributed, instead being clustered somewhat around the center of the hard, perturbative
jet.  This points to the possibility of fitting for the exponent in the UE ansatz of Eq.~\ref{acfwUE}, but we do not explore this possibility any further.

The average UE densities $\bar \Lambda_{\rm UE}$ defined from the extracted curves $\Lambda_{\rm UE}(R)$ are listed in 
Table \ref{UEtable}.  These are defined by averaging $\Lambda_{\rm UE}(R)$ in the range from $R=0.4$ to $R=0.8$, 
with the perturbative contribution
to $\langle \Delta {\cal G}(R)\rangle$ matched at $R_{\rm min}=0.25$.  Also listed are the average transverse momentum densities 
$\Lambda_{\rm Trans}$ in the transverse region (for the definition of which see Sec.~\ref{trans}).  For all the samples we see that
$\bar \Lambda_{\rm UE}$ tracks $\Lambda_{\rm Trans}$ as expected, although it tends to be somewhat larger.  For the three Pythia8
samples we see how, as we scale the amount of UE activity, $\Lambda_{\rm Trans}$ and $\bar \Lambda_{\rm UE}$ rise linearly and in tandem.

 We can ask how sensitive the extraction of $\bar \Lambda_{\rm UE}$ is to deviations from the flatness assumption.
The largest dependence on the perturbative ansatz $C$ in Eq.~\ref{lambda} is through the term $\langle \Delta {\cal G}(R)\rangle-C$.  
A large overestimation (underestimation) of $C$ gives a large underestimation (overestimation) of the average UE density.  For example, if
for Pythia8 Tune 4C we let $C$ be given by the $\langle \Delta {\cal G}(R)\rangle$ obtained from Pythia8 in the absence of UE/ISR, we find an
${\cal O}(100\%)$ correction to $\bar \Lambda_{\rm UE}$.  In contrast the corresponding correction for Herwig++ is only ${\cal O}(20\%)$.
  
\begin{table}
\begin{center}
\begin{tabular}{l | c | c}
Monte Carlo Sample & $\Lambda_{\rm Trans}$ & $\bar \Lambda_{\rm UE}$ \\
\hline
\hline 
Pythia8 Tune 4C& 3.2 GeV& $2.8\pm 0.3$ GeV\\
Pythia8 Tune 4$\rm{C}^{\prime}$ & 4.6 GeV  & $5.0\pm 0.7$ GeV \\
Pythia8 Tune 4$\rm{C}^{\prime\prime}$ & 6.0 GeV & $7.2 \pm 1.1$ GeV  \\
Herwig++ Tune LHC-UE7-2 & 3.3 GeV & $3.7\pm1.0$ GeV\\
Herwig++ Tune LHC-UE7-2 ($R^{3.3}$ ansatz) & 3.3 GeV & $3.2\pm0.4$ GeV
\end{tabular}
\caption{Table of extracted UE densities.  $\Lambda_{\rm Trans}$ is the average $p_T$ density in the 
transverse region and $\bar \Lambda_{\rm UE}$ is extracted from $\langle \Delta{\cal G}(R)\rangle$.  The `error' bars quoted
for $\bar \Lambda_{\rm UE}$ are the maximum difference of the function $\Lambda_{\rm UE}(R)$ from its average, which
is computed in the range from $R=0.4$ to $R=0.8$.  Tunes 4$\rm{C}^{\prime}$ and 4$\rm{C}^{\prime\prime}$ differ from tune 4C in that
they have twice and thrice as much UE activity, respectively. See Appendix \ref{mc} for more details about the Monte Carlo.}\label{UEtable}
 \end{center}
 \end{table}

The validity of the procedure explored in this section depends on a number of assumptions, and an assessment of its usefulness would require
a detailed experimental study with a particular emphasis on the systematic uncertainties involved.  In addition, a better theoretical
understanding of the perturbative contribution to $\langle \Delta{\cal G}(R)\rangle$ would be needed to determine the degree to which
the flatness assumption is warranted. Nevertheless, as
the luminosity (and eventually the center of mass energy) continues to rise at the LHC, 
the increasing importance of UE and PU motivates the study of additional experimental handles on the
impact of UE and PU on hadronic final state reconstruction.  While something like the ``jet-area/median'' method introduced
in Ref.~\cite{arXiv:0912.4926} is much more ambitious, since it provides a handle on fluctuations in UE and PU, the procedure proposed here
might be a useful complement to this and other methods.

\section{Angular correlations in the transverse region}\label{trans}

To gain more intuition for the scaling information encoded in the average angular structure
function, it is interesting to consider a qualitatively different ensemble of jets.  
Unlike many traditional jet shape observables, $\langle \Delta{\cal G}(R)\rangle$ is a meaningful observable for regions of the detector 
populated with soft, diffuse radiation.  In the following we will focus on the transverse region of 
dijet events, as traditionally employed in underlying event studies.  We will explore three different models for the underlying event:
(i) the analytically tractable Feynman-Wilson gas; (ii) a toy Monte Carlo that describes uniformly distributed mini-jets; and (iii) full Pythia8 simulation.  

\begin{figure}
\centering
    \includegraphics[width=5.1cm]{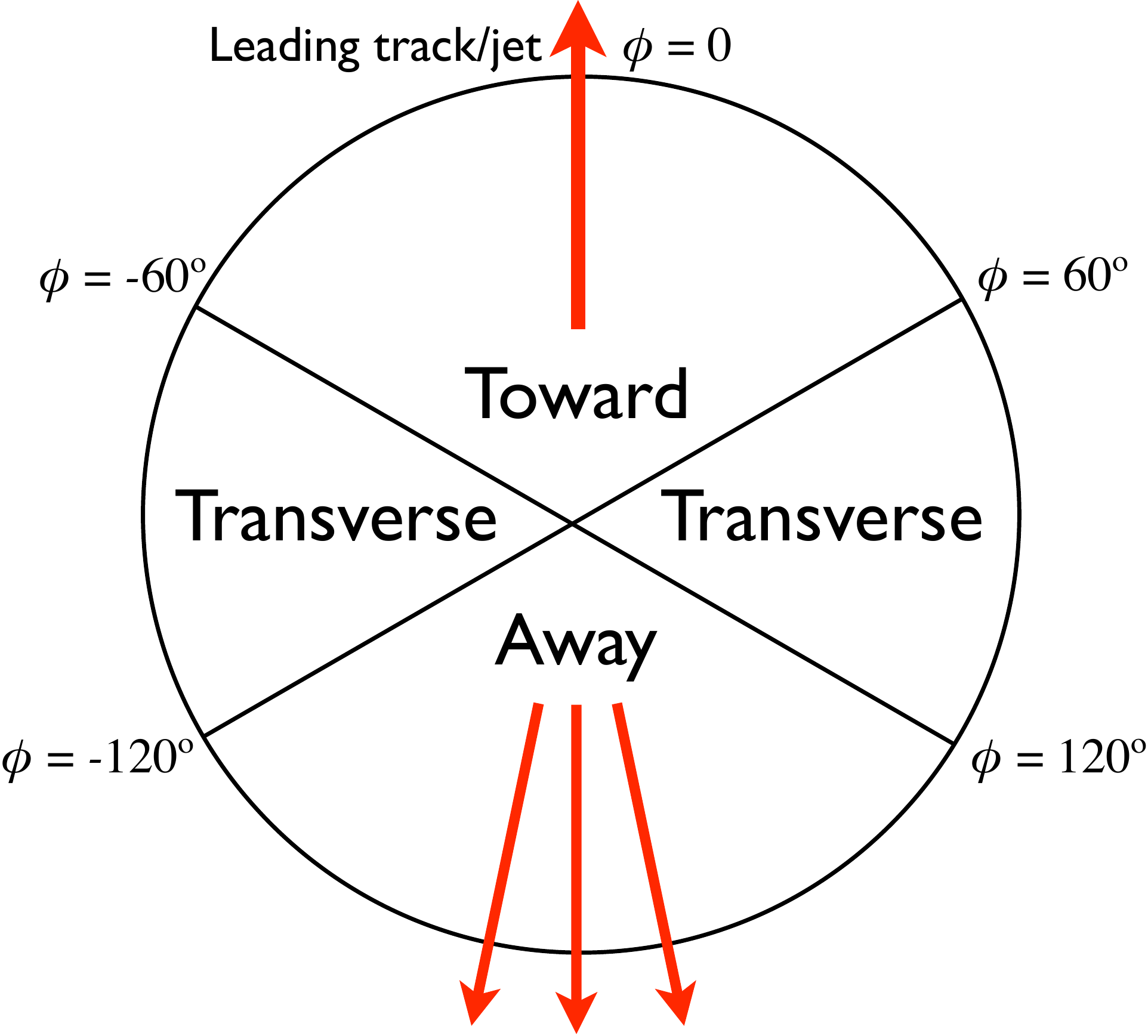}
\caption{Definition of the transverse region.}\label{transdef}
\end{figure}

The transverse region is defined with respect to the direction of the hardest jet or hardest
charged track ($\phi\equiv0$), with its two halves extending between $\phi=\pm\pi/3$ and $\phi=\pm2\pi/3$ and $\eta=-\eta_{\rm{max}}$ 
and $\eta=\eta_{\rm{max}}$ (see Fig.~\ref{transdef}) \cite{Affolder:2001xt}.
We choose to define the transverse region with respect to the hardest jet and set $\eta_{\rm{max}}=2.0$. 
The two transverse regions are further
distinguished by their total activity.  The region with the largest scalar sum $p_T$ is denoted as
the trans-max region and the other is the trans-min region.  Historically, underlying event studies have often focused 
on these two regions, since they are less sensitive to contamination from the hard perturbative scattering \cite{Field:2011iq}.
In the following ensemble averages will always be defined on charged tracks in the trans-min region, although other variations are possible.

The most commonly used models of the underlying event are based on a picture of hadronic collisions in which
multiple parton interactions (MPI) contribute to the soft activity in each event \cite{LU-TP-87-5}.  
These models proceed by extrapolating the 2 $\rightarrow$ 2 QCD matrix element to semi-perturbative energies.  Spectator
partons that are not involved in the hard collision are allowed to scatter, resulting in an approximately uniform
population of low $p_T$ jets, at least up to moderate pseudorapidities.  

\subsection{Feynman-Wilson Gas}

Following the MPI hypothesis, we will make a number of simplifying
assumptions and see what sort of $\langle \Delta{\cal G}(R)\rangle$ we would expect to measure in the
trans-min region. To begin with consider a model for the
underlying event along the lines of the Feynman-Wilson gas \cite{Feynman:1969ej}.
We assume that the number of underlying event particles is distributed with a probability distribution $\rho(n)$ that depends
only on the number of particles $n$.  Typically this is taken to
be Poissonian.  Also, we will assume that the transverse momentum of the 
underlying event particles has a distribution $p_T(n)$ that is uniform in $\eta$ and $\phi$.

In this case the average angular correlation function can be calculated from
\beq
\langle{\cal G}(R)\rangle=\sum_{n=2}^\infty \rho(n) \sum_{i<j\leq n} p_T(n)p_T(n)\Delta R^2_{ij}\Theta_{\rm dR}(R-\Delta R_{ij}) 
\eeq{unacftrans}
For fixed $n$, the sum over the pairs of constituents breaks up into ${n \choose 2}$ identical
terms so that we can write
\beq
\langle{\cal G}(R)\rangle=\sum_{n=2}^\infty {n \choose 2} \rho(n) p_T(n)p_T(n) 
\int [d\eta d\phi] \  \Delta R_{12}^2\Theta_{\rm dR}(R-\Delta R_{12}) 
\eeq{unacftrans1}
where $[d\eta d\phi] \equiv d\eta_1 d\eta_2 d\phi_1 d\phi_2$.
Since all the $R$ dependence is in the integral, taking the logarithmic derivative yields the simple expression:
\beq
\langle \Delta {\cal G}(R) \rangle \equiv {d \log \langle{\cal G}(R)\rangle \over d\log R} = R{\int [d\eta d\phi] \  \Delta R_{12}^2\delta_{\rm dR}(R-\Delta R_{12}) \over \int [d\eta d\phi] \  \Delta R_{12}^2\Theta_{\rm dR}(R-\Delta R_{12})} 
\eeq{asftrans}
That is, assuming that the underlying event is uniformly distributed, 
we can compute $\langle \Delta {\cal G}(R) \rangle$ without knowing the $p_T$ or particle number distributions.
The average angular structure function becomes purely geometric.

For a rectangle of size $\Delta \eta \times \Delta \phi$, $\langle \Delta{\cal G}(R) \rangle$
can be computed for $dR=0$.  For example, for $R < \min(\Delta \eta, \Delta \phi)$ we have:
\beq
\langle\Delta{\cal G}(R)\rangle =\frac{ 2\pi\Delta\eta \Delta\phi  - 4(\Delta\eta+ \Delta\phi) R + 2R^2}{ {\pi \over 2}\Delta\eta \Delta\phi  - {4 \over 5}(\Delta\eta+ \Delta\phi) R + {1 \over 3}R^2} = 4 -\frac{8}{5\pi}\frac{\Delta \eta +\Delta \phi}{\Delta \eta \Delta \phi} R + {\cal O}(R^2)\
\eeq{envelope}
The corresponding
$\langle\Delta{\cal G}(R)\rangle$ over the full range of $R$ with $dR=0.04$, $\Delta \phi=\pi/3$, and $\Delta \eta=4.0$ is plotted as the black curve
in Fig.~\ref{transmmc}. 
Edge effects quickly bring $\langle\Delta{\cal G}(R)\rangle$ down to zero for $R\gtrsim1$.  At small but nonzero $R$, $\langle\Delta{\cal G}(R)\rangle$ approaches the value of 4 that is obtained in the infinite-plane
limit $\Delta \eta, \Delta \phi \to \infty$.  As $R\to0$ the smoothing pulls $\langle\Delta{\cal G}(R)\rangle$ down from near 4, as would be
the case for $dR=0$ (see Eq.~\ref{envelope}).  

\subsection{Toy Monte Carlo}\label{transmc}

\begin{figure}
\centering
    \includegraphics[width=9.4cm]{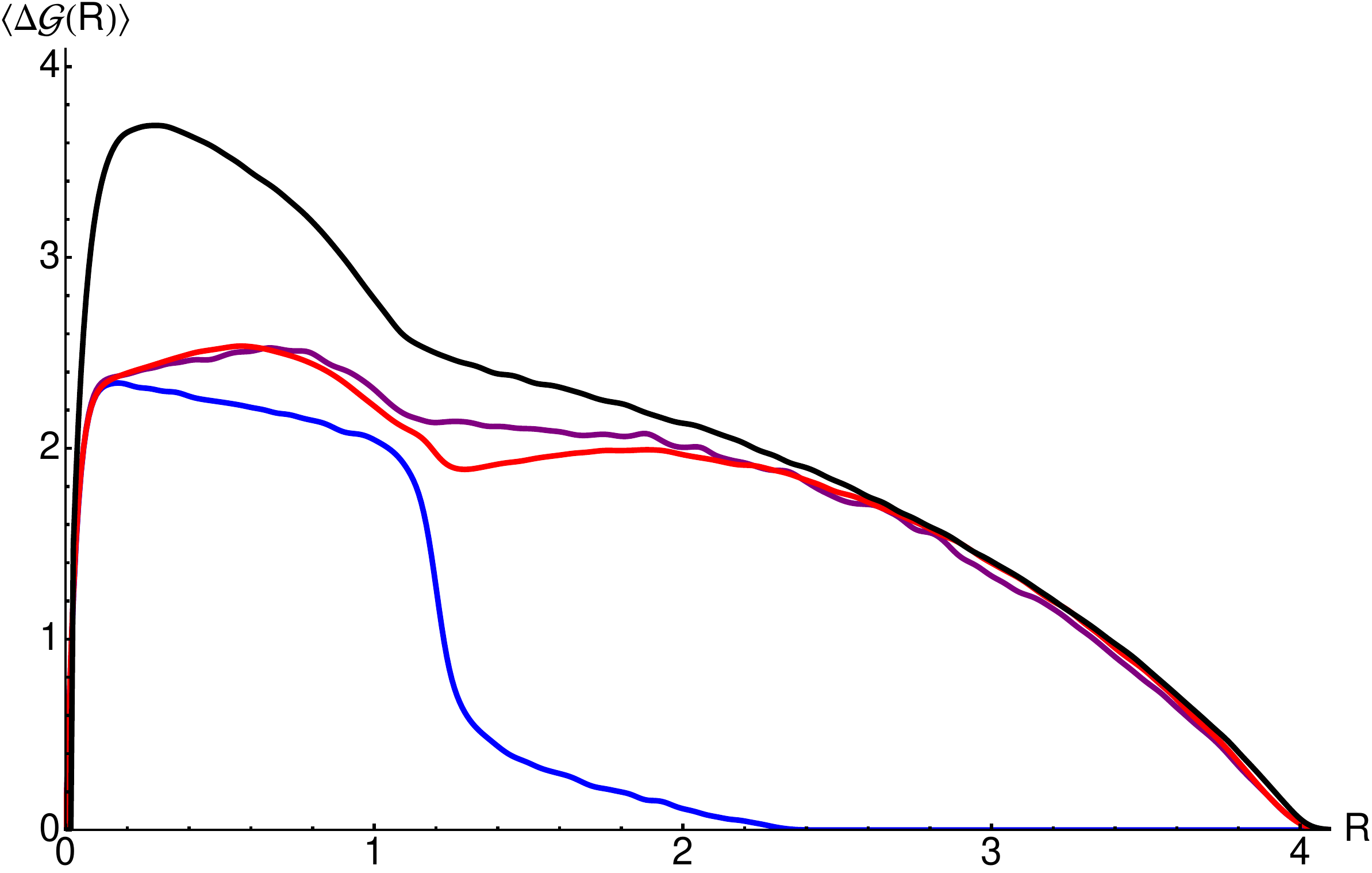}
\caption{$\langle\Delta{\cal G}(R)\rangle$ in the trans-min region. The black curve corresponds to the Feynman-Wilson gas.  The purple curve
corresponds to Pythia8 dijet events with $p_{T {\rm jet}}>200$ GeV.  The red curve corresponds to the toy Monte Carlo, described below.  Finally, the blue curve corresponds to an ensemble
of single ``DLA mini-jets'' as employed in the toy Monte Carlo.  See Appendix \ref{mc} for more details about the Monte Carlo.}\label{transmmc}
\end{figure}
With Monte Carlo we can explore the $\langle\Delta{\cal G}(R)\rangle$ that results from underlying event
models more complicated than the simplest Feynman-Wilson gas.  In particular
we would like to understand the $\langle\Delta{\cal G}(R)\rangle$ that results from MPI models. 
\begin{figure}
\centering
    \includegraphics[width=9.4cm]{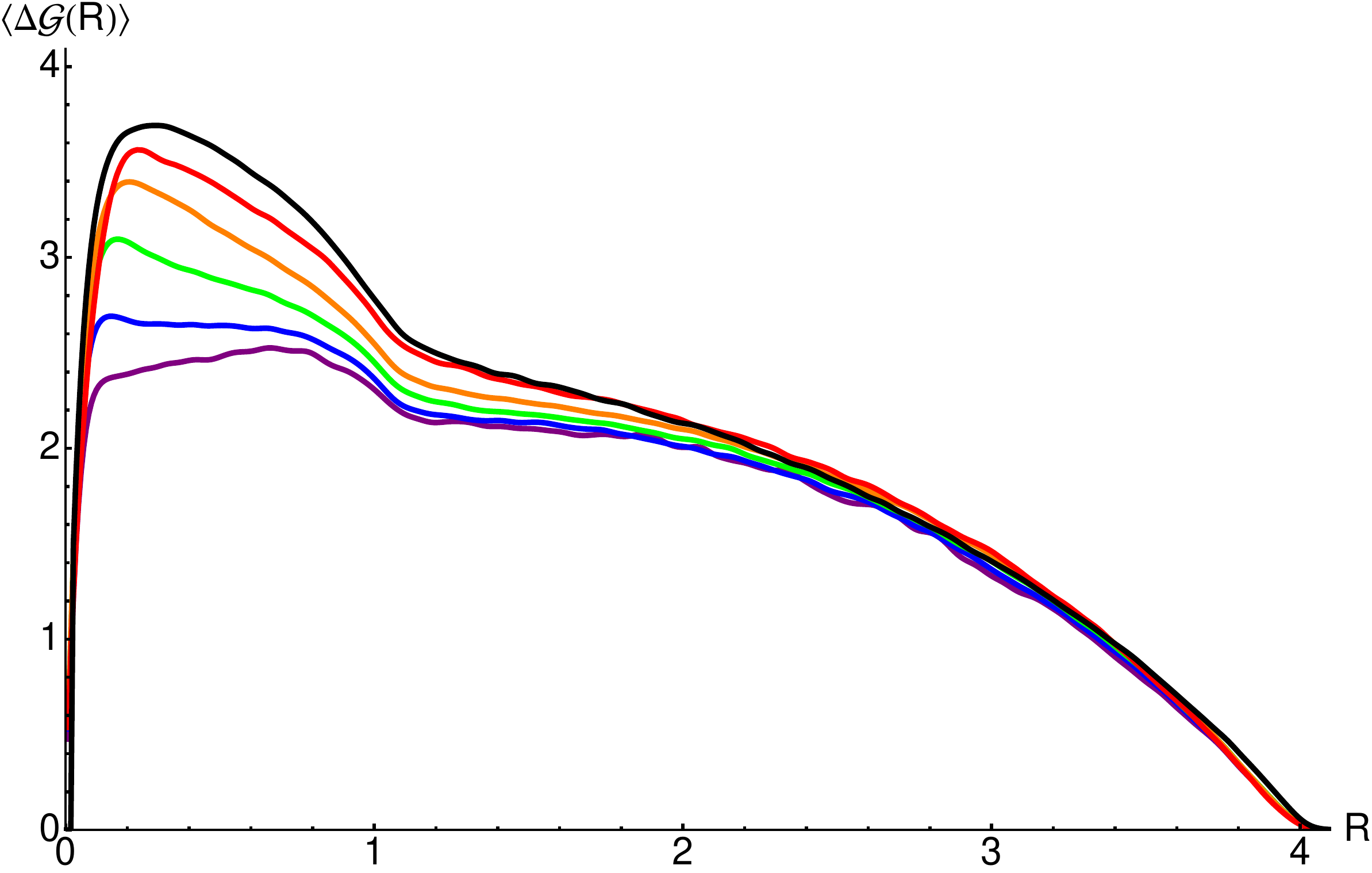}
\caption{$\langle\Delta{\cal G}(R)\rangle$ for ensembles excluding events where the $p_T$ of the hardest charged particle in the trans-min region exceeds
a given $p_{T {\rm max}}$.  
$\langle \Delta {\cal G}(R)\rangle$ for the Feynman-Wilson gas is shown in black, while the colored curves correspond to Pythia8 dijet events with $p_{T {\rm jet}}>200$ GeV.  
The purple curve has no $p_T$ cut, while the blue, green, orange and red curves correspond to $p_{T {\rm max}}$ of 16 GeV, 8 GeV, 4 GeV and 1 GeV, respectively.  
See Appendix \ref{mc} for more details about the Monte Carlo.}\label{ptcut}
\end{figure}
We use the following toy Monte Carlo to test whether we understand the key physics ingredients entering
into the form of $\langle\Delta{\cal G}(R)\rangle$ obtained from Pythia8.  First, we generate $n$ mini-jets, where $n$ is 
selected from a Poisson distribution with mean $\langle n \rangle = 0.30$ per unit area of azimuth/pseudorapidity.
 We choose the transverse momenta of the mini-jets from the distribution
\beq
\rho(p_T) \sim \frac{dp_T^2}{{p_T^4}}\Theta(p_T^2-\lambda^2)
\eeq{ptdistro}
where $\lambda=1$ GeV.  To generate the emissions that make up
each mini-jet we employ a sort of double logarithmic approximation (DLA) 
scheme that mimics the parton shower.  That is, for each mini-jet we generate a fixed number
of emissions, $N_g=10$, where each ``gluon" is generated according to the distribution
\beq
w(\theta, z, \phi) \sim \frac{d\theta}{\theta} \frac{dz}{z} \frac{d\phi}{2\pi}
\eeq{emissiondistro}
where $\theta$ is the angular separation between the gluon and the center of the mini-jet, $z$
is the energy fraction of the gluon, and $\phi$ is the azimuthal angle around the
mini-jet axis.  The $\theta$ and $z$ distributions are cut off at finite values, with $0.1<z<1.0$
and $0.01 < \theta < 1.2$.  

The $\langle\Delta{\cal G}(R)\rangle$ for an ensemble of single ``DLA mini-jets'' is plotted as the
blue curve in Fig.~\ref{transmmc}.  By construction, for $R < \theta_{\rm max}=1.2$ DLA mini-jets have
an average angular structure function near the perturbative value of 2.  
Since MPI models are dominated by two scaling behaviors, (i) the perturbative
$R^2$ scaling of the substructure of the mini-jets at small $R$ and (ii) the Feynman-Wilson
scaling at large $R$, we expect our toy model to yield a $\langle\Delta{\cal G}(R)\rangle$ between the limiting
$\langle\Delta{\cal G}(R)\rangle$ curves of (i) and (ii).  This is what we see in Fig.~\ref{transmmc}, where
the red curve interpolates between the blue and black curves, with a smooth
transition between the two regimes at intermediate $R$.  Furthermore, we see that the toy Monte Carlo
does a good job of describing the average angular structure function obtained from Pythia8.

\subsection{Emergence of jets}\label{jetemerge}
In Fig.~\ref{ptcut} we illustrate the effect on $\langle\Delta{\cal G}(R)\rangle$ when the ensemble is defined to exclude events where the $p_T$ of the 
hardest charged particle in the trans-min region exceeds a given $p_{T {\rm max}}$. The Feynman-Wilson curve is shown for comparison in black.  As the 
cut is decreased from infinity, $\langle \Delta {\cal G}(R)\rangle$ approaches the Feynman-Wilson curve at small $R$.  
This is because the $p_T$ cut has the effect of removing events with harder (and therefore jettier) structure in the transverse
region.  The soft events that remain are composed of soft mini-jets whose substructure is less jetty due to
the smaller dynamical range of the parton shower.  This effect should be measurable in data.

\section{Discussion and conclusions}\label{discuss}

The angular correlation function provides a direct probe into the scaling behavior of QCD.  Because it is
formulated in terms of two-particle correlations, it asks a particularly detailed question about jet substructure.  
Many jet shape observables (the integrated jet shape, angularities, etc.) are linear in the momenta of the
jet constituents and are explicitly defined with respect to the jet center, i.e.~they are radial moments of one
kind or another.  In this sense they can be thought of as accessing one-particle correlations, and the
measurement of $\langle \Delta {\cal G}(R)\rangle$ would be expected to provide orthogonal information about the
substructure of jets.  

This comment is especially relevant to the observed differences in $\langle \Delta {\cal G}(R)\rangle$ between Herwig++
and Pythia8.  If these event generators are being tuned to experimental data for observables
like the integrated jet shape, then it is not surprising that they should give different predictions for two-particle correlations.  
Understanding these differences will be important if measurements of $\langle \Delta {\cal G}(R)\rangle$ 
are to be used for improving Monte Carlo tunes. 
For example, the evolution variable of the parton shower 
affects both the structure of emissions in the jet as well as the scale at which the running coupling is evaluated.  In addition, numerous other effects
such as hadronization, color reconnections, and the details of the UE model will enter into the prediction for $\langle \Delta {\cal G}(R)\rangle$.
One would also like to understand how matched samples of jets, with their different treatment of hard, wide-angle radiation, affect the
form of $\langle \Delta {\cal G}(R)\rangle$.   
 
An especially interesting property of $\langle \Delta {\cal G}(R)\rangle$ is that it can be interpreted as an average scaling exponent.
This makes the leading order result $\langle \Delta {\cal G}(R)\rangle=2$ particularly simple to understand and provides
useful intuition for the higher order effects explored in Sec.~\ref{analytics}.  It also forms the basis of the two applications of 
$\langle \Delta {\cal G}(R)\rangle$ examined in Sec.~\ref{ueinjet} and Sec.~\ref{trans}.  
In the first case, reliable extraction of $\bar \Lambda_{\rm UE}$ will require
a better theoretical understanding of the perturbative contributions to $\langle \Delta {\cal G}(R)\rangle$.  In particular it 
remains to determine to what degree the flatness assumption made in Sec.~\ref{proc} is warranted.  In the second case,
forming the ensemble average in the transverse region provides additional insight into the physics encoded in $\langle \Delta {\cal G}(R)\rangle$.
We find that MPI models predict a quasi-universal form for $\langle \Delta {\cal G}(R)\rangle$ with $\langle \Delta {\cal G}(R)\rangle \simeq2$
at small $R$ and the large $R$ form following the Feynman-Wilson gas.  Although the jetty nature of UE is already well established,
a measurement of $\langle \Delta {\cal G}(R)\rangle$ in the transverse region has the nice property that it exhibits both the perturbative substructure and the uniform distribution of mini-jets.

In this paper we have argued that the ensemble average of the angular structure function makes for a particularly interesting jet shape observable.
From a theoretical point of view its interpretation as a scaling exponent is especially compelling.  From an experimental
point of view the possibility of measuring the average contribution of the underlying event and pile-up to hard perturbative jets
is intriguing.  To go further will require experimental input, and we hope that $\langle \Delta {\cal G}(R)\rangle$ might be measured at the LHC.

\Acknowledgements

We thank Michael Peskin for innumerable discussions on QCD and jets and for his comments on the draft of this paper.
We thank Michael Spannowsky for helpful comments on the draft of this paper.
We thank Ariel Schwarzman and Peter Loch for many useful conversations about jet measurements in ATLAS.  
We thank Spencer Gessner for his work on implementing the angular correlation function in the ATLAS analysis framework.  
We thank Jon Walsh for stimulating discussions on factorization and SCET.
We thank Michael Seymour and Peter Skands for very helpful discussions of the underlying event models in Herwig++ and Pythia8.  
We thank Michael Seymour and the University of Manchester Particle Theory Group for their hospitality and support during a recent visit.
We thank the Galileo Galilei Institute for Theoretical Physics for their hospitality and the INFN for support throughout the `Interpreting LHC Discoveries' workshop.  
This work is supported by the US Department of Energy under contract DE--AC02--76SF00515.  
M.J.~receives partial support from the Stanford Institute for Theoretical Physics.  
A.L.~is supported in part by the U.S. National Science Foundation, grant NSF-PHY-0969510, the LHC Theory Initiative, Jonathan Bagger, PI.

\appendix
\section{Monte Carlo}\label{mc}
Here we give details of how the various Monte Carlo samples were generated.  Throughout Herwig++ refers to Herwig++ v.~2.5.1 
\cite{Bahr:2008pv,Bahr:2008dy,Bahr:2009ek,Gieseke:2011na}, and
Pythia8 refers to Pythia8 v.~8.150 \cite{Sjostrand:2006za,Sjostrand:2007gs,Corke:2010yf}.  
To cluster jets, we use the FastJet v.~2.4.2 \cite{Cacciari:2011ma} implementation of the anti-kT algorithm \cite{Cacciari:2008gp}.
Note that the final state shower in Herwig++ is angular-ordered, while Pythia8 has a $p_T$-ordered shower.
We simulate $pp$ collisions at a center of mass energy of $E_{\rm CM} = 7$ TeV.
No attempt is made to model detector effects, with particle-level information being used in all cases.  
In Sec.~\ref{analytics}, the samples have been generated with MPI and ISR turned off.  In Herwig++ this is done via the flags
\begin{verbatim}
    ShowerHandler:MPIHandler NULL
    SplittingGenerator:ISR No
\end{verbatim}  
while in Pythia8 it is accomplished via the flags
\begin{verbatim}
    PartonLevel:MI = off
    SpaceShower:QCDshower = off
\end{verbatim}
Also, in Sec.~\ref{analytics} we use ensembles composed of either quark or gluon jets.  These are obtained by choosing the hardest jet
in each event from samples generated with pure quark or glue final states, i.e.~$gg\to q\bar{q}$ and $q\bar{q}\to q\bar{q}$ in the case
of quark jets and $gg\to gg$ and $q\bar{q} \to gg$ in the case of gluon jets.  For the event samples in Sec.~\ref{ueinjet} MPI and ISR are again
turned on.  For the underlying event Herwig++ makes use of tune `LHC-UE7-2' \cite{herwigtune} and Pythia8 makes use of 
tune `4C' \cite{Corke:2010yf,Buckley:2011ms}.  In all cases we have checked that the $p_T$ distributions
are similar between corresponding Pythia8 and Herwig++ samples.
Also, in Sec.~\ref{ueinjet} we
use two Pythia8 event samples (tunes 4$\rm{C}^{\prime}$ and 4$\rm{C}^{\prime\prime}$) 
in which MPI activity has been increased by factors of $2.0$ and $3.0$.  This is done by dialing
the parameter {\tt MultipleInteractions:Kfactor}.  

Finally, in Sec.~\ref{trans} the ensemble averages over the trans-min region come
from a dijet sample where the $p_T$ of the leading jet is greater than 200 GeV.  The dijet sample is generated with Pythia8 using tune 4C,
and the ensemble averages only make use of charged particles.  Because of the large angular separations between charged tracks
that can occur in the trans-min region, the ensemble average is modified to include a $R_{\rm min}$ prescription.  That is, a given
event only contributes to the numerator and denominator of $\langle \Delta {\cal G}(R)\rangle$ in Eq.~\ref{avedg} for $R \ge R_{\rm min}$, where $R_{\rm min}$
is the minimal angular separation between charged tracks in the trans-min region of that particular event.  


\end{document}

%% file: mymacros.tex
\def\beq{\begin{equation}}
\def\eeq#1{\label{#1}\end{equation}}
\def\eeqn{\end{equation}}

\newenvironment{Eqnarray}%
   {\arraycolsep 0.14em\begin{eqnarray}}{\end{eqnarray}}
\def\beqa{\begin{Eqnarray}}
\def\eeqa#1{\label{#1}\end{Eqnarray}}
\def\eeqan{\end{Eqnarray}}
\def\CR{\nonumber \\ }

\let\bar=\overbar

\def\lsim{\mathrel{\raise.3ex\hbox{$<$\kern-.75em\lower1ex\hbox{$\sim$}}}}
\def\gsim{\mathrel{\raise.3ex\hbox{$>$\kern-.75em\lower1ex\hbox{$\sim$}}}}

\def\del{\partial}
\def\Dslash{\not{\hbox{\kern-4pt $D$}}}
\def\dslash{\not{\hbox{\kern-2pt $\del$}}}
\def\pslash{\not{\hbox{\kern-2pt $p$}}}
\def\qslash{\not{\hbox{\kern-2pt $q$}}}
\def\kslash{\not{\hbox{\kern-2pt $k$}}}
\def\oneslash{\not{\hbox{\kern-2pt $1$}}}
\def\twoslash{\not{\hbox{\kern-2pt $2$}}}
\def\threeslash{\not{\hbox{\kern-2pt $3$}}}
\def\fourslash{\not{\hbox{\kern-2pt $4$}}}
\def\aslash{\not{\hbox{\kern-2pt $a$}}}
\def\rslash{\not{\hbox{\kern-2pt $r$}}}
\def\Jslash{\not{\hbox{\kern-2pt $J$}}}

\def\msb{{\bar{\scriptsize M \kern -1pt S}}}

\def\drb{{\bar{\scriptsize D \kern -1pt R}}}

\def\apb#1 {  \langle #1 ] }

\makeatletter
\def\section{\@startsection{section}{0}{\z@}{5.5ex plus .5ex minus
 1.5ex}{2.3ex plus .2ex}{\large\bf}}
\def\subsection{\@startsection{subsection}{1}{\z@}{3.5ex plus .5ex minus
 1.5ex}{1.3ex plus .2ex}{\normalsize\bf}}
\def\subsubsection{\@startsection{subsubsection}{2}{\z@}{-3.5ex plus
-1ex minus  -.2ex}{2.3ex plus .2ex}{\normalsize\sl}}

\renewcommand{\@makecaption}[2]{%
   \vskip 10pt
   \setbox\@tempboxa\hbox{\small #1: #2}
   \ifdim \wd\@tempboxa >\hsize     
       \small #1: #2\par          
     \else                        
       \hbox to\hsize{\hfil\box\@tempboxa\hfil}
   \fi}